\documentclass[prx, twocolumn,longbibliography]{revtex4-2}
 
\usepackage{float}
\usepackage{bbm}
\usepackage{epsfig}
\usepackage{epstopdf}
\usepackage{graphicx, subfigure}
\usepackage{amsmath,amssymb}
\usepackage{color}
\usepackage[colorlinks=true,citecolor=blue,linkcolor=magenta]{hyperref}
\usepackage[capitalize]{cleveref}
\usepackage{braket,mleftright}

\usepackage{dsfont}
\usepackage{extarrows}
\usepackage{physics}
\UseRawInputEncoding 

\newcommand{\rev}[1]{\textcolor{black}{#1}}

\begin{document}

\title{Fast high-fidelity single-qubit gates for flip-flop qubits in silicon}

\author{F.~A.~Calderon-Vargas}
\altaffiliation{Present address: Sandia National Laboratories, Livermore, CA 94550, USA}
\author{Edwin~Barnes}
\author{Sophia~E.~Economou}
\affiliation{ Department of Physics, Virginia Tech, Blacksburg, Virginia 24061, USA}

\begin{abstract}
The flip-flop qubit, encoded in the states with antiparallel donor-bound electron and donor nuclear spins in silicon, showcases long coherence times, good controllability, and, in contrast to other donor-spin-based schemes, long-distance coupling. Electron spin control near the interface, however, is likely to shorten the relaxation time by many orders of magnitude, reducing the overall qubit quality factor. Here, we theoretically study the multilevel system that is formed by the interacting electron and nuclear spins and derive analytical effective two-level Hamiltonians with and without periodic driving. We then propose an optimal control scheme that produces fast and robust single-qubit gates in the presence of low-frequency noise without relying on parametrically restrictive sweet spots. This scheme increases considerably both the relaxation time and the qubit quality factor. 
\end{abstract}
\maketitle

\section{Introduction}\label{sec: Intro}

Quantum computation promises to revolutionize the scientific world, from fundamental science to information technology~\cite{Nielsen2010}.
In the ongoing race to build the first fully operational quantum computer, donor spin qubits in isotopically purified silicon ($^{28}\mathrm{Si}$)~\cite{Itoh2014} are promising candidates due to their long coherence times and their integrability with metal-oxide-semiconductor structures~\cite{Witzel2010,Tyryshkin2012,Zwanenburg2013,Harvey-Collard2017a,Chatterjee2020,Struck2020,Morello2020}. Donor spins present coherence times reaching around half a minute (half a second) for the nuclear (electron) spin~\cite{Muhonen2014,Tenberg2018}, up to hours in bulk ensembles~\cite{Saeedi2013}, and a high degree of controllability~\cite{Pla2013,Muhonen2015,Muhonen2018}. However, the implementation of two-qubit gates has proven to be quite challenging. Most of the approaches for two-qubit operations are based on Kane's seminal proposal~\cite{Kane1998}, where the qubit coupling is achieved via the exchange interaction between donor-bound electrons. The use of such short-range interactions requires near-atomic precision in the placement of the donors~\cite{Dehollain2014,Song2016d}. And, even though recent works have shown more relaxed requirements on the precision of donor placement~\cite{Kalra2014,Hill2015,Madzik2020}, long-distance coupling is still challenging without inserting intermediate couplers~\cite{Trifunovic2012,Mohiyaddin2016,Pica2016}.

A recent proposal by Tosi et al.~\cite{Tosi2017b} circumvents the precise donor placement limitation by using the electric dipole, created when the electron is shared between the donor and the Si/SiO$_2$ interface, as a long range coupling between a pair of qubits, each encoded in the flip-flop states of the donor-bound electron and donor nuclear spins. These qubits, hereafter called \textit{flip-flop qubits}, can be fully controlled by microwave electric fields through hyperfine modulation. A constant (dc) electric field induces qubit rotations about the $z$-axis, while an oscillating (ac) electric field implements $x$ and $y$ gates. The gate control of the electron near the interface, however, may cause flip-flop relaxation via spontaneous phonon emission, as shown in Ref.~\onlinecite{Boross2016}, resulting in a relaxation time $T_1$ approximately 8 orders of magnitude shorter than in bulk~\cite{Pines_1957}, and a few orders of magnitude shorter than the $T_1$ predicted in Ref.~\onlinecite{Tosi2017b}. This lowers the qubit quality factor ($T_1/\tau$ with $\tau$ being the qubit gate time), which gives the number of available qubit operations before coherence is lost. A high quality factor is one of the main requirements for fault-tolerant quantum computing~\cite{Fowler2012}. One way of improving the quality factor would be to increase $T_1$ by reducing the external magnetic field $B_0$~\cite{Tosi2017b}. However, the magnetic field strength used in the experiments is usually between 0.4 $T$ and 1.4 $T$. \rev{This is because the qubit readout via spin-dependent tunneling~\cite{morello_single-shot_2010}  requires the qubit Zeeman splitting  to be larger than $\sim5 k_B T_{e}$, which is the thermal broadening of the electron reservoir at temperature $T_e$ (usually between 100 mK and 200 mK)~\cite{Dehollain2013}.} Therefore, lowering the magnetic field strength is not desirable. Another approach would be to use optimal control pulses that implement faster qubit gates in the magnetic field range used in the laboratory , which is the approach we take here.

In this paper, we propose optimally designed control pulses for fast high-fidelity single-qubit gates, i.e. arbitrary $z$- and $x$-rotations, for flip-flop qubits. We use both time-independent and time-dependent Schrieffer-Wolff transformations~\cite{Schrieffer1966} to derive effective qubit Hamiltonians for both ac and dc driving. The former, required to implement $x$-rotations, is studied in the strong driving regime using Floquet perturbation theory~\cite{Shirley1965}. With the analytical effective qubit Hamiltonian, we are able to produce single-qubit gates that are much faster and more robust than previous proposals, with fidelities above $99.99\%$. Moreover, our scheme does not rely on restricting parameters to operational sweet spots, like clock transitions~\cite{Tosi2017b}, allowing us, for example, to find fast gates for different magnetic fields strengths, increasing the relaxation time and the qubit quality factor considerably.

The paper is organized as follows. In Sec.~\ref{sec: Spin_and_orbital_Hamiltonians}, we analyze the flip-flop qubit system and derive a simplified Hamiltonian in the combined spin and orbital eigenbases. Then, in Sec.~\ref{sec:Rz_gates}, we introduce an effective qubit Hamiltonian with no oscillating driving and use it to produce fast high-fidelity $z$-rotations. In Sec.~\ref{sec:Rx_gates}, we use time-dependent Schrieffer-Wolff perturbation theory (which is discussed in detail in App.~\ref{appendix:SW transformation}) to derive an effective two-level Hamiltonian with oscillating driving and, via Floquet perturbation theory, present analytical expressions for the resonance and Rabi frequencies, which are used to produce fast high-fidelity $x-$rotations. We conclude in Sec.~\ref{sec: Conclusions}.

\section{Spin and orbital Hamiltonians}\label{sec: Spin_and_orbital_Hamiltonians}
\begin{figure}[tb]
 \includegraphics[width=0.7\linewidth]{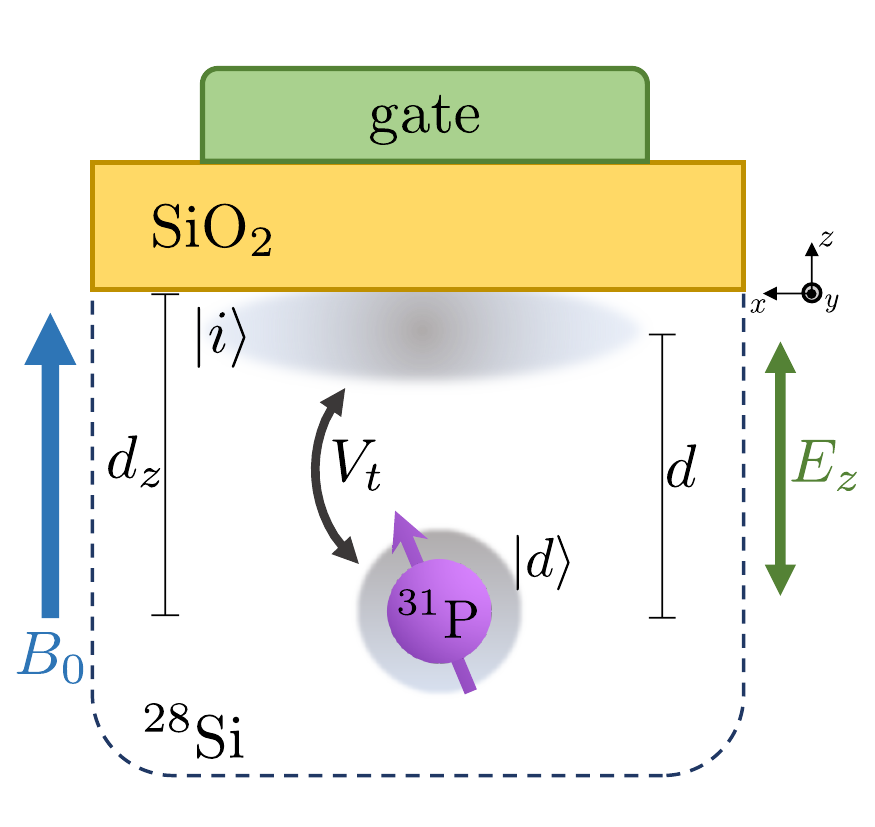}
 \caption{Schematic of the flip-flop qubit system. A phosphorus donor is embedded in $^{28}$Si at a depth $d_z$ from the interface with a thin SiO$_2$ layer. A top metal gate controls the position of the wavefunction of the donor unpaired electron via an electric field $E_z$. The electron orbit is quantized into a $\ket{d}$ state at the donor and a $\ket{i}$ state at the interface. The wavefunctions of these states are schematically shown in gray.
 }
 \label{fig:system_schematic}
 \end{figure}

The setup of the system follows the experimental proposal from Ref.~\cite{Tosi2017b}, where the wavefunction of the donor-bound electron of a phosphorus donor ($^{31}$P) embedded in isotopically purified $^{28}$Si is controlled by a vertical electric field $E_z$ applied by a metal gate on top (Fig.~\ref{fig:system_schematic}). The donor is at a depth $d_z$ from the interface with a thin SiO$_2$ layer. The electron (nuclear) spin $S=1/2$ ($I=1/2$) has a gyromagnetic ratio $\gamma_e/2\pi=27.97~\text{GHz}/\text{T}$ ($\gamma_n/2\pi=17.23~ \text{MHz}/\text{T}$) and basis states $\{\ket{\uparrow},\ket{\downarrow}\}$ ($\{\ket{\Uparrow},\ket{\Downarrow}\}$).
For an isolated $^{31}$P donor atom in unstrained Si, the isotropic Fermi-contact hyperfine interaction $A$ in the non-relativistic limit is proportional to the probability amplitude $|\psi (0,0,d_z)|^2$ of the unpaired electron wavefunction at the nucleus. 
Under a large magnetic field $B_0$ ($B_0(\gamma _e+\gamma _n)\gg A$) along the $z$-axis, the spin Hamiltonian is
\begin{equation}\label{eq:Spin_Hamiltonian}
H_{spin}=\gamma_e B_0 S_z-\gamma_n B_0 I_z+A\vec{S}\cdot\vec{I},
\end{equation}
where $S_z=\tfrac{\hbar}{2}(\ket{\uparrow}\bra{\uparrow}-\ket{\downarrow}\bra{\downarrow})$ ($I_z=\tfrac{\hbar}{2}(\ket{\Uparrow}\bra{\Uparrow}-\ket{\Downarrow}\bra{\Downarrow})$) is the $z$ component of the electron (nuclear) spin operator. The flip-flop states $\{\ket{\uparrow\Downarrow} ,\ket{\downarrow\Uparrow}\}$ are effectively decoupled from the other states. Consequently, our analysis is henceforth focused on the Hilbert space spanned by the flip-flop states, but all the numerical results reported in this work are obtained keeping the full Hilbert space.

The shifting of the electron wavefunction by the electric field $E_z$ creates an electric dipole $\mu_d=d\,e$, where $e$ is the electron charge and $d$ is the distance between the center-of-mass positions of the donor-bound ($\ket{d}$) and interface-bound ($\ket{i}$) orbitals (see Fig.~\ref{fig:system_schematic}). Following Ref.~\cite{Tosi2017b}, we can use these two well-defined positions as two orthogonal quantum states, and describe the electron orbital dynamics with a simple two-level Hamiltonian:
\begin{equation}\label{eq:Orbital_Hamiltonian}
H_{orb}=-\frac{d\,e\,\Delta E}{2\hbar}\tau_z^{id}+\frac{V_t}{2}\tau_x^{id},
\end{equation}
where $\Delta E=E_z-E_0$ is the deviation of the vertical electric field away from the ionization point $E_0$ (electric field value where the stationary electron charge is equally distributed between $\ket{i}$ and $\ket{d}$), $V_t$ is the tunnel coupling between the orbital states $\ket{i}$ and $\ket{d}$, and $\tau_z^{id}=\ket{i}\bra{i}-\ket{d}\bra{d}$, $\tau_x^{id}=\ket{i}\bra{d}+\ket{d}\bra{i}$ are Pauli operators. \rev{In this work, the qubit is operated with electric field $E_z$ near the donor ionization field $E_0$ or $E_z<E_0$, where the valley splitting between the lower $\ket{i}$ and upper $\ket{v}$ valley interface states is much larger than the tunnel coupling $V_t$~\cite{Tosi2017b} and far from the anticrossing between $\ket{d}$ and $\ket{v}$. }

Given the hyperfine dependence on the electron position, the hyperfine interaction $A(E)$ changes from the bulk value $A/2\pi \approx $ 117 MHz to $A \approx $ 0 when the electron is fully displaced to the interface. The electron gyromagnetic ratio also differs from its value at the donor when the electron is confined at the Si/SiO$_2$ interface; the difference $\Delta_{\gamma}$ can be up to 0.7\%~\cite{Rahman2009a,Tosi2017b}. The orbital position dependence of these two energies can be incorporated in the Hamiltonian by treating them as projection operators in the orbital Hilbert space, i.e. $A(\ket{d}\bra{d})$ and $\gamma_e B_0\Delta_{\gamma}\ket{i}\bra{i}$. 

The total Hamiltonian combines the spin and orbital degrees of freedom and, in the basis $\{\ket{g\otimes\!\uparrow\Downarrow},\ket{g\otimes\!\downarrow\Uparrow},\ket{e\otimes\!\uparrow\Downarrow},\ket{e\otimes\!\downarrow\Uparrow}\}$, has the following form:
\begin{equation}\label{eq:full_Hamiltonian}
\begin{aligned}
H=&-\frac{\varepsilon_0}{2}\sigma_{zi}+\frac{\varepsilon_z}{2}\sigma_{iz}+\frac{\Delta_{\varepsilon_z}}{4}\left(\sigma_{iz}+\sin\theta\,\sigma_{xz}+\cos\theta\,\sigma_{zz}\right)\\
&+\frac{A}{8}\left(2\sigma_{ix}-2\sin\theta\,\sigma_{xx}-2\cos\theta\,\sigma_{zx}\right.\\
&\left.+\sin\theta\,\sigma_{xi}+\cos\theta\,\sigma_{zi}\right),
\end{aligned}
\end{equation}
where $\ket{g}$ ($\ket{e}$) is the ground (excited) eigenstate of the orbital Hamiltonian~\eqref{eq:Orbital_Hamiltonian}, $\varepsilon_0=\sqrt{(d\,e\,\Delta E/\hbar)^2+V_t^2}$ and $\varepsilon_z=B_0(\gamma _e+\gamma _n)$ are the orbital and Zeeman energy splittings, $\Delta_{\varepsilon_z}=B_0\gamma_e\Delta_{\gamma}$ is the Zeeman energy shift when the electron is at the interface, $\tan\theta=V_t\hbar/(d\,e\,\Delta E)$ is a mixing angle, and $\sigma_{pq}=\tau_p\otimes \zeta_q$ is the Kronecker product of Pauli operators with $\sigma_{zz}=(\ket{g}\bra{g}-\ket{e}\bra{e})\otimes(\ket{\uparrow\Downarrow}\bra{\uparrow\Downarrow}-\ket{\downarrow\Uparrow}\bra{\downarrow\Uparrow})$.

The Hamiltonian acquires a simpler form in the basis formed by the orbital and hyperfine eigenstates, $\{\ket*{g~\widetilde{\uparrow\Downarrow}},\ket*{g~\widetilde{\downarrow\Uparrow}},\ket*{e~\widetilde{\uparrow\Downarrow}},\ket*{e~\widetilde{\downarrow\Uparrow}}\}$. In this basis, the spin energy splitting is conditioned on the orbital state and is given by $\varepsilon_{s^{(\mp)}}=\sqrt{(A(1\mp\cos\theta)/2)^2+(\varepsilon_z+\Delta_{\varepsilon_z}(1\pm\cos\theta)/2)^2}$, where $(-)$ and $(+)$ corresponds to the orbital ground and excited eigenstates, respectively. Owing to the orbital ($\varepsilon_0$), Zeeman ($\varepsilon_z$), and spin energy splittings being much larger than the hyperfine interaction ($A$) and Zeeman energy shift ($\Delta_{\varepsilon_z}$), we neglect terms of the form $\eta/\varepsilon$ where $\eta\in \{A,\Delta_{\varepsilon_z}\}$ and $\varepsilon\in\{\varepsilon_0,\varepsilon_z,\varepsilon_{s^{(\mp)}}\}$. The simplified Hamiltonian in the basis  $\{\ket*{g~\widetilde{\uparrow\Downarrow}},\ket*{g~\widetilde{\downarrow\Uparrow}},\ket*{e~\widetilde{\uparrow\Downarrow}},\ket*{e~\widetilde{\downarrow\Uparrow}}\}$ is 
\begin{widetext}
\begin{equation}\label{eq:H_matrix_form_spin_orbit_eigenbases}
\tilde{H}=\left(
\begin{array}{cccc}
-\frac{\varepsilon_0-\varepsilon_{s^{(-)}}}{2}+\frac{A\cos\theta}{8} & 
 0 & 
 \frac{(A+2\Delta_{\varepsilon_z})\sin\theta}{8} & 
 -\frac{A\sin\theta}{4} \\
 0 & 
 -\frac{\varepsilon_0+\varepsilon_{s^{(-)}}}{2}+\frac{A\cos\theta}{8} & 
 -\frac{A\sin\theta}{4} & 
 \frac{(A-2\Delta_{\varepsilon_z})\sin\theta}{8} \\
 \frac{(A+2\Delta_{\varepsilon_z})\sin\theta}{8} &
 -\frac{A\sin\theta}{4} & 
 \frac{\varepsilon_0+\varepsilon_{s^{(+)}}}{2}-\frac{A\cos\theta}{8} & 
 0 \\
 -\frac{A\sin\theta}{4} & 
 \frac{(A-2\Delta_{\varepsilon_z})\sin\theta}{8} & 
 0 & 
 \frac{\varepsilon_0-\varepsilon_{s^{(+)}}}{2}-\frac{A\cos\theta}{8}\\
\end{array}
\right).
\end{equation}
\end{widetext}
 
  \begin{figure}[tbp]
 \includegraphics[width=0.95\linewidth]{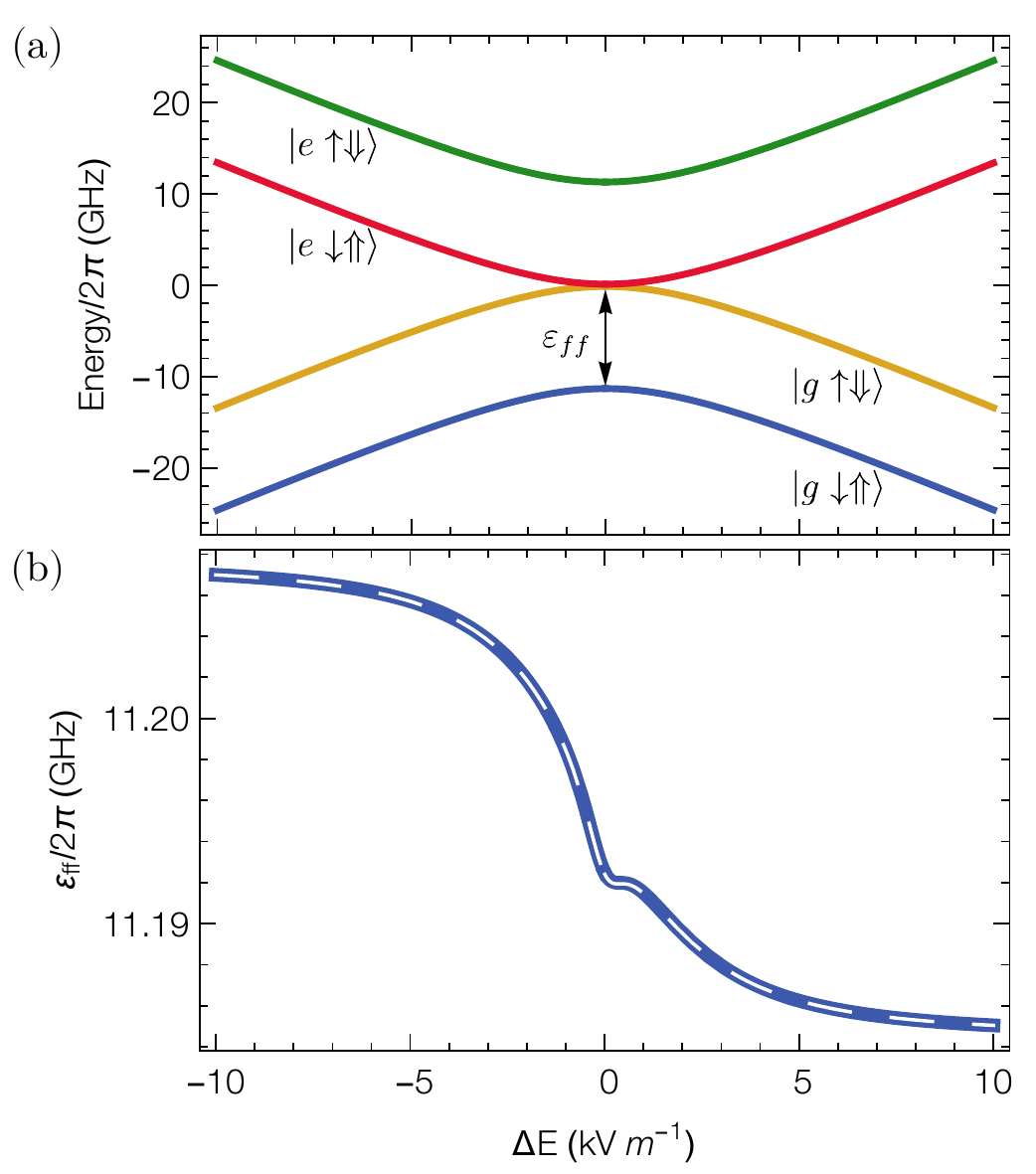}
\caption{\textbf{(a)} Energy-level diagram of the flip-flop system. The flip-flop qubit is encoded in the two lowest-energy eigenstates, with $\varepsilon_{ff}$ being the transition energy between the two qubit levels. Moreover, note that for $E_z \gg E_0$ ($\Delta E\gg 0$) the qubit states $\ket{0}$ and $\ket{1}$ are effectively equal to the states $\ket{g\downarrow\Uparrow}$ and $\ket{g\uparrow\Downarrow}$, respectively. Similarly, for $E_z \ll E_0$ ($\Delta E\ll 0$) the excited levels are effectively equal to $\ket{e\downarrow\Uparrow}$ and $\ket{e\uparrow\Downarrow}$. \textbf{(b)} The numerically calculated transition energy $\varepsilon_{ff}$ (blue curve), where the kink in the curve near $\Delta E=0$ is the clock transition (reduced dephasing). The white dashed curve overlaying the blue one is obtained with the analytical expression for the effective flip-flop transition energy $\varepsilon_{ff}$, Eq.~\eqref{eq:effective_flip-flop_transition}, showing excellent agreement with the numerical result.
The system parameters used to plot both figures are $B_0=0.4~\text{T}$, $V_t/2\pi=11.44~\text{GHz}$, $d=15~\text{nm}$, and $\Delta_{\gamma}=-0.002$. 
 }
 \label{fig:energy and transition frequency}
 \end{figure}
 
The flip-flop qubit is encoded in the two lowest-energy eigenstates of the total Hamiltonian, which are approximately $\ket*{g~\widetilde{\downarrow\Uparrow}}$ and $\ket*{g~\widetilde{\uparrow\Downarrow}}$. The eigenenergies of the system Hamiltonian are shown in Fig.~\hyperref[fig:energy and transition frequency]{\ref*{fig:energy and transition frequency}(a)}. Note that the qubit states $\{\ket{0},\ket{1}\}$ are effectively $\{\ket{g\downarrow\Uparrow},\ket{g\uparrow\Downarrow}\}$ for $\Delta E\gg0$ ($\theta\approx 0$), which corresponds to fully displacing the electron to the interface. Moreover, at $\abs{\Delta E}\gg 0$ ($\theta\approx\{0,\pi\}$) the electron is fully displaced  either to the interface or to the donor, and as the flip-flop qubit is effectively decoupled from electric fields, these are referred to as idling regions. Conversely, the electron must be displaced to the region around the ionization point $\Delta E=0$ ($\theta=\pi/2$) in order to implement any quantum gate. This is also the region, however, where the qubit is most sensitive to electrical noise and leakage. The latter can be reduced by applying slow-varying pulses to retain adiabaticity. The main source of noise in this type of system is charge noise, usually stemming from defects and electron traps at the Si/SiO$_2$ interface. Given that the qubit gates for donor qubits takes less than a microsecond, the charge noise is usually static within a single gate and, therefore, we can model it as quasi-static noise. For this noise, Ref.~\cite{Tosi2017b} shows the presence of ``clock transitions'' in the flip-flop transition energy, i.e. regions where the transition is noise-insensitive up to a certain order. This is clearly shown in Fig.~\hyperref[fig:energy and transition frequency]{\ref*{fig:energy and transition frequency}(b)}, where, for a specific set of parameters, a second-order clock transition is found at $\Delta E\approx 0.4~\text{kV~m}^{-1}$. However, as we will show in the following sections, it is possible to implement robust rotations that do not use the clock transition as an operating point. This can soften experimental requirements and improve the quality of gates at the same time.

Donor spin qubits are among the most coherent solid state quantum systems, and the flip-flop qubit is not expected to be an exception~\cite{Tosi2017b}. Nonetheless, a theoretical description ~\cite{Boross2016} of the phonon-mediated relaxation of the flip-flop qubit shows that when the electron is at the ionization point ($\Delta E=0$), the flip-flop relaxation time $T_1$ is a few orders of magnitude shorter than what Ref.~\onlinecite{Tosi2017b} predicts and around 8 orders of magnitude shorter than what was predicted for a P donor in bulk silicon~\cite{Pines_1957}. This can be counteracted, however, by increasing the tunnel coupling $V_t$ which, according to Tosi et al. proposal~\cite{Tosi2017b}, should be able to be tuned by at least two orders of magnitude. Evidently, the ratio $T_1^{(i)}/T_1^{(j)}$ ($i,j$ referring to two different sets of system parameters) given by~\cite{Boross2016}
\begin{equation}\label{eq:T_1_ratio_relaxation_time}
    \frac{T_1^{(i)}}{T_1^{(j)}}=\left[\frac{\varepsilon_0^2(\varepsilon_0^2-(\gamma_e B_0)^2)^2}{V_t^4(\gamma_e B_0)^3}\right]^{(i)} \left[\frac{V_t^4(\gamma_e B_0)^3}{\varepsilon_0^2(\varepsilon_0^2-(\gamma_e B_0)^2)^2}\right]^{(j)},
\end{equation}
with $\varepsilon_0=\sqrt{(d\,e\,\Delta E/\hbar)^2+V_t^2}$, shows that increasing the tunnel coupling in ($i$) relative to ($j$) and keeping the other parameters equal does indeed increase the relaxation time of ($i$) with respect to ($j$). In the following sections, we show that it is possible to implement fast high-fidelity single-qubit gates with different magnetic field strengths and tunnel coupling values, improving the qubit quality factor.

For the analysis and results reported in this work, unless stated otherwise, we use the same parameters reported in Ref.~\cite{Tosi2017b}. Accordingly, the distance $d$ is equal to $15~\text{nm}$, and $\Delta_{\gamma}=-0.002$.

\section{ $R_z(\phi)$ gates and effective Hamiltonian without oscillating driving}\label{sec:Rz_gates}

Qubit rotations about the $z$-axis in the flip-flop system are implemented by displacing the electron from an idling point $\Delta E_{idle}$ (preferably near the interface where the hyperfine interaction is effectively null) toward an operating point $\Delta E_{op}$ in the region around the ionization point, parking there for a certain amount of time, and then returning to the initial point $\Delta E_{idle}$. We consider two operating points, one at the clock transition $\Delta E=0.4~\text{kV~m}^{-1}$ (as proposed in Ref.~\onlinecite{Tosi2017b}), and the other beyond the ionization point and closer to the donor ($\Delta E=-12~\text{kV~m}^{-1}$), both under the same magnetic field strength, $B_0=0.4$~T, and  tunnel coupling, $V_t/2\pi=11.44~\text{GHz}$. We also consider larger magnetic fields and larger tunnel couplings, $\{B_0=0.8~\text{T},V_t/2\pi=22.55~\text{GHz}\}$ and $\{B_0=1.2~\text{T},V_t/2\pi=33.71~\text{GHz}\}$, both with operating points closer to the donor, $\Delta E=-20~\text{kV~m}^{-1}$ and $\Delta E=-30~\text{kV~m}^{-1}$, respectively. The operating points closer to the donor produce high-fidelity gates but are not unique: any operating point closer to the donor could also produce high-fidelity gates. This is because in that region the flip-flop qubit dephasing rate is much lower than near the ionization point. As shown in Fig.~\ref{fig:pulse,fidelity many rotations}(b), the dephasing rate in the region closer donor is as low as or lower than the dephasing rate at the second order clock transition. Now, if $\Delta E_{op}$ is too close to the ionization point ($\Delta E=0$), the applied electric field must vary slowly when approaching the fast dephasing region around the ionization point to preserve adiabaticity and avoid leakage to unwanted excited states.
This can be accomplished with a smooth pulse, a modified `Planck-taper' window function~\cite{McKechan2010}:
\begin{equation}\label{eq:smoothed_square_pulse}
\Xi(t) = \xi_0+\begin{cases}
\frac{\xi_f-\xi_0}{1+\exp\left[\frac{t_r}{t}+\frac{t_r}{t-t_r}\right]/\varsigma} & t_0 < t < t_r, \\
\xi_f-\xi_0 & t_r \leq t \leq (T - t_r), \\
\frac{\xi_f-\xi_0}{1+\exp\left[\frac{-t_r}{t-T+t_r}-\frac{t_r}{t-T}\right]/\varsigma} & (T - t_r) < t < T, \\
0 & t \leq t_0 \text{ or } t \geq T.
   \end{cases}
\end{equation}
Here, $t_0$ is the time at the start of the pulse, $t_r$ is the ramp time, $T$ is the gate time ($T>2t_r$), $\xi_0$ is the value of the control field at the start and end of the pulse, $\xi_f$ is the control field value at the pulse plateau, and $\varsigma> 0$ modulates the pulse slope such that for $\varsigma>1$ it is decreased in the region between the pulse inflection points and plateau, see Fig.~\hyperref[fig:pulse,fidelity many rotations]{\ref*{fig:pulse,fidelity many rotations}(a)}. The latter is useful for preserving the adiabaticity when the pulse plateau is in a region in close proximity to excited states. For the control of the donor electron position, we use $\xi_0=\Delta E_{idle}>>0$, such that the electron is at or near the interface, and $\xi_f=\Delta E_{op}$ is the electric field magnitude that places the electron at the operating point.

 \begin{figure}[tb]
 \includegraphics[width=\linewidth]{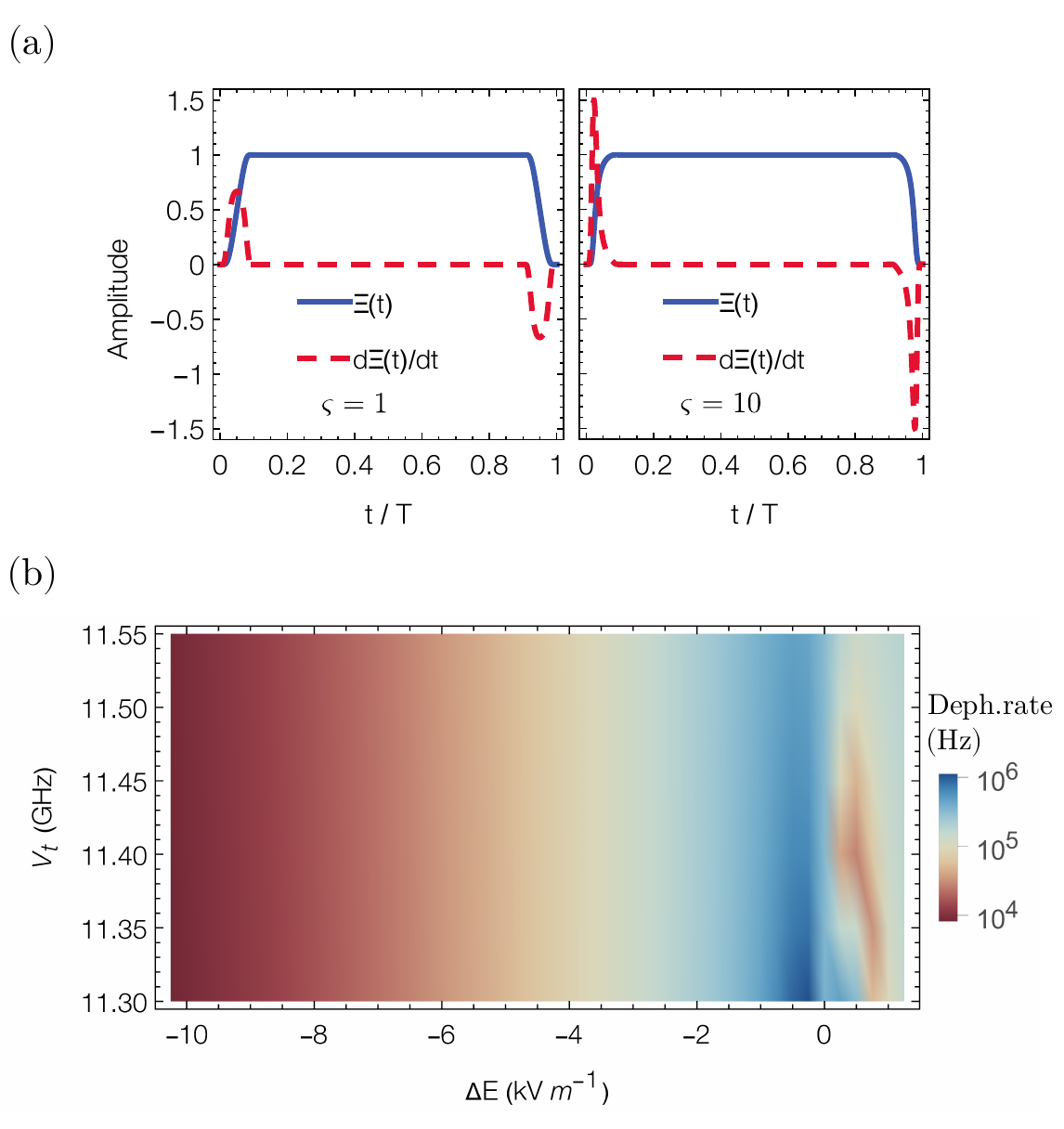}
\caption{\textbf{(a)} Example of a pulse using Eq.~\eqref{eq:smoothed_square_pulse} with arbitrary parameters. The pulses (solid blue curves) in both panels have the same ramp time $t_r/T=0.1$ (and same gate time $T$) but different value for $\varsigma$. Note that for larger $\varsigma$ the pulse slope of the region between the inflection point and the plateau is lowered, which is evidenced by the pulse first derivative curve (dashed red line). The pulse first derivative curves in both panels are reduced in amplitude for visualization purpose.
\textbf{(b)} Estimated flip-flop qubit dephasing rate, assuming electric field noise $\delta E_{z,\text{rms}} =100~\text{V m}^{−1}$. In the region around $\Delta E=10~\text{kV m}^{-1}$ and also at $\Delta E\approx 0.4~\text{kV m}^{-1}$ (clock transition), the qubit dephasing rate is at least two orders of magnitude smaller than near the ionization point $\Delta E=0$.
 }
 \label{fig:pulse,fidelity many rotations}
 \end{figure}

The amount of time the electron should remain parked at the operating point to implement some desired qubit rotation can be determined with an analytic effective Hamiltonian in the qubit logical space. Noting that the off-diagonal elements of the Hamiltonian~\eqref{eq:H_matrix_form_spin_orbit_eigenbases} are smaller than the diagonal ones, we use a time-independent Schrieffer-Wolff (SW) transformation~\cite{Schrieffer1966} (a.k.a. van Vleck or quasi-degenerate perturbation theory~\cite{VanVleck1929,Shavitt1980,Winkler2003}, see Appendix~\ref{appendix:SW transformation}) up to fourth order to diagonalize the Hamiltonian~\eqref{eq:H_matrix_form_spin_orbit_eigenbases}. We consider up to fourth order because we will need an expression for the transition energy as precise as possible to find the optimum conditions to generate high-fidelity $R_x(\phi)$ rotations with an oscillating magnetic field. Note that the off-diagonal elements are non-negligible only near the ionization point $\Delta E=0$ ($\theta=\pi/2$), and in this region the orbital-conditioned spin energy splittings are effectively equal ($\varepsilon_{s^{(-)}}\approx\varepsilon_{s^{(+)}}$), which is  assumed in the SW transformation. The resulting effective Hamiltonian in the qubit space is $H_{ff}=-\tfrac{1}{2}\varepsilon_{ff}\sigma_z$ (with $\sigma_z=\ket{0}\bra{0}-\ket{1}\bra{1}$) and the flip-flop transition energy $\varepsilon_{ff}$ is  given by
\begin{equation}\label{eq:effective_flip-flop_transition}
\begin{aligned}
\varepsilon_{ff}=&\varepsilon_{s^{(-)}}-A\sin^2\theta\left( \frac{\Delta_{\varepsilon_z}}{8\varepsilon_0}+\frac{A\varepsilon_{s^{(-)}}}{8\Delta_{os}^2}\right.\\
&\left. \frac{A^2\varepsilon_0\varepsilon_{s^{(-)}}\cos\theta}{16\Delta_{os}^4}-\frac{A^3\varepsilon_0^2\varepsilon_{s^{(-)}}\sin^2\theta}{32\Delta_{os}^6}
\right),
\end{aligned}
\end{equation}
where $\Delta_{os}^2\equiv\varepsilon_0^2-\varepsilon_{s^{(-)}}^2$. Figure~\hyperref[fig:energy and transition frequency]{\ref*{fig:energy and transition frequency}(b)} shows excellent agreement between the analytical expression for the flip-flop transition and the numerical result.

The magnitude of $\varepsilon_{ff}/2\pi$ in the idling region ($\Delta E\gg1$) is on the order of GHz, therefore, in order to have an identity operation when the electron is at the idling point $\Delta E_{idle}$ we move to a frame rotating with a frequency equal to the flip-flop qubit precession frequency at the idling point. Therefore, the evolution operator in this frame is $\tilde{U}(t,t_0)=U_0^{\dagger}(t,t_0)U(t,t_0)$. Here, $U(t,t_0)=\mathcal{T}\{\exp\text{(}-i\int_{t_0}^t H(\tau) \dd{\tau} \text{)} \}$ is the evolution operator with the time-dependent Hamiltonian given by Eq.~\eqref{eq:full_Hamiltonian}, and $U_0(t,t_0)=\exp\left(-i H_0 (t-t_0)\right)$ is the evolution operator with the time-independent Hamiltonian $H_0\equiv H(\Delta E_{idle})=-\tfrac{1}{2}\varepsilon_{s^{(-)}}(\Delta E_{idle})\sigma_z$, where the constant $\Delta E_{idle}$ is chosen such that $\theta\approx 0$ in \eqref{eq:full_Hamiltonian}. Then the donor electron displacement from the idling point to the operating point implements a rotation about the $z$-axis with an angle (phase accumulated) given by
\begin{equation}\label{eq:phase_accumulated_integral}
\phi=-\int_{t_0}^t(\varepsilon_{ff}(\tau)-\varepsilon_{s^{(-)}}(\Delta E_{idle}))\dd{\tau}.
\end{equation}

\begin{figure}[tbp]
	\includegraphics[width=1\linewidth]{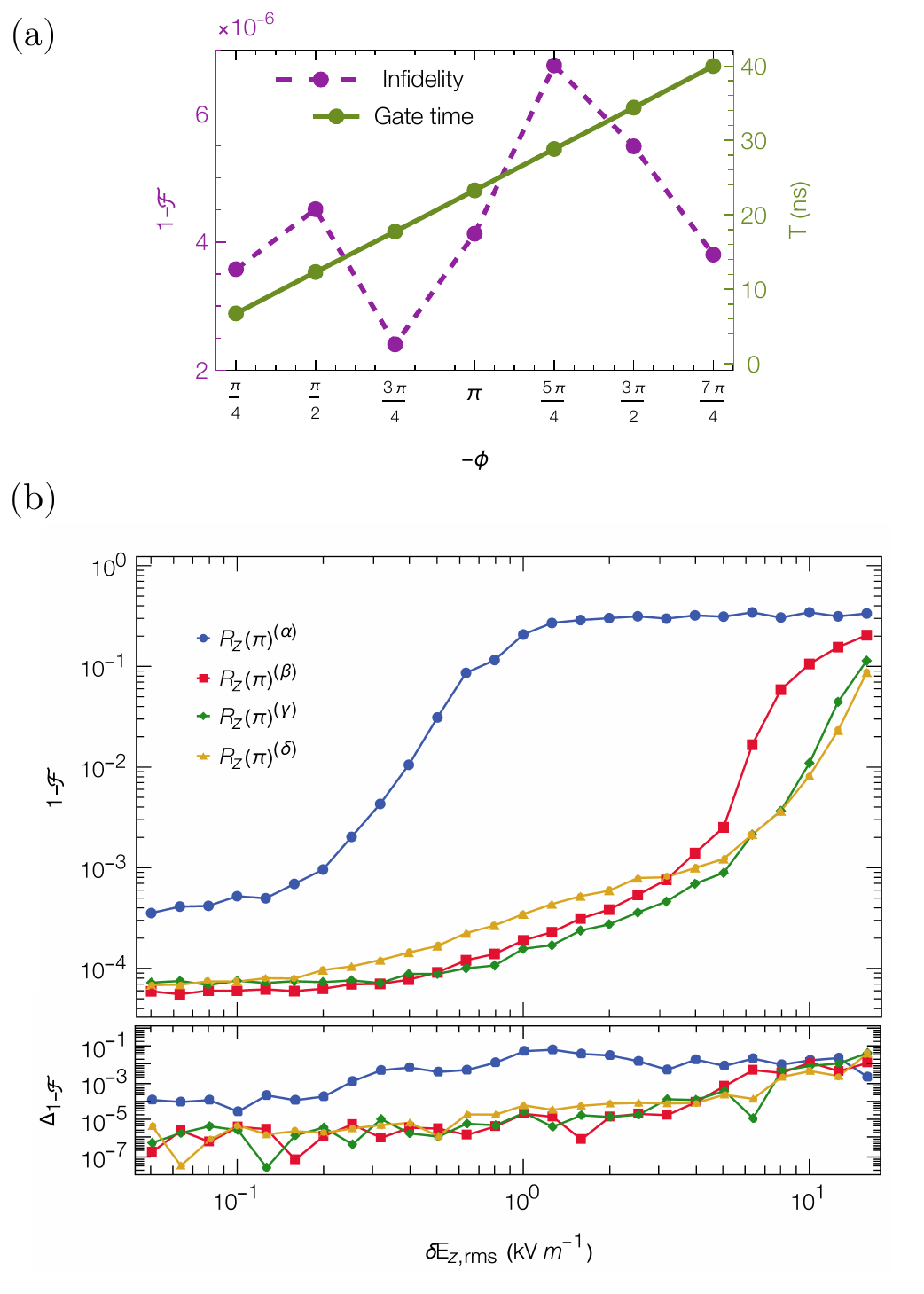}
\caption{\textbf{(a)} Infidelity, $1-\mathcal{F}$~\eqref{eq:fidelity_equation}, in the absence of noise, and gate times for $z$-rotations about different angles. The pulse parameters used are $\xi_0=60~\text{kV~m}^{-1}$, $\xi_f=-12~\text{kV~m}^{-1}$, $t_r=0.9~\text{ns}$, and $\varsigma=1$. The infidelity for each rotation is on the order of $10^{-6}$, and the relative variation between rotation infidelities is solely due to fluctuations in the convergence of the numerical search algorithm for increasing integration times. Finally, note that $R_z(-\phi)\equiv \exp[i \phi\sigma_z/2]$ is equivalent to $e^{i\pi}R_z(\varphi)\equiv e^{i\pi}\exp[-i\varphi\sigma_z/2]$, where $\varphi\equiv 2\pi -\phi$.
\rev{\textbf{(b)} (Upper part) Infidelity of $\pi$ $z$-rotations as a function of various electric field noise strength values $\delta E_{z,\text{rms}}$ and fixed tunnel coupling noise $\delta V_{t,\text{rms}}$. The tunnel coupling noise and pulse parameters for each gate are given in Table~\ref{tab:table_1}. (Lower part) Change in infidelity $\Delta_{1-\mathcal{F}}$ between gate infidelity obtained with only electric field noise $\delta E_z$ and gate infidelity obtained with noise on both electric field $\delta E_z$ and tunnel coupling $\delta V_t$.}
}
\label{fig:pi fidelity, other fidelity, filter function}
\end{figure}

\begin{table}[htb]
    \centering 
    \caption{System and pulse parameters for each infidelity curve in Fig.~\hyperref[fig:pi fidelity, other fidelity, filter function]{\ref*{fig:pi fidelity, other fidelity, filter function}(b)}. \rev{ For each curve, the stated average gate fidelity $\mathcal{F}$ is calculated  assuming  electric field noise  $\delta E_{z,\text{rms}}=100~\mathrm{V~m}^{-1}$ and tunnel coupling noise $\delta V_{t,\text{rms}}$ indicated in this table.}}
    \vspace{1mm} 
\begin{tabular}{ |p{2.4cm}|| p{1.1cm}|p{1.1cm}|p{1.1cm}| p{1.1cm}| }
 \hline
   & $R_z(\pi)^{(\alpha)}$ & $R_z(\pi)^{(\beta)}$ & $R_z(\pi)^{(\gamma)}$ & $R_z(\pi)^{(\delta)}$\\
 \hline
 $B_0$~(T)   & 0.4    & 0.4 &  0.8 &  1.2\\
$ V_t/2\pi$~(GHz)   & 11.44    & 11.44 &  22.55 &  33.71\\
$ \Delta E_{op}$~(kVm$^{-1}$)   & 0.4    & -12 &  -20 &  -30\\
$ t_r$    & 4.3    & 0.9 &  0.9 &  0.8\\
$ \varsigma$    & 70   & 1 &  1 &  1\\
$\delta V_{t,\text{rms}}/2\pi $~(MHz)		 & 2.7 & 2.7 & 3.3 & 3.5 \\
$ T$~(ns)   & 70.35   & 23 &  12.36 &  24.04\\
$ \mathcal{F}$~(\%)  & 99.95  & 99.994 &  99.992 &  99.993\\
 \hline
\end{tabular}\label{tab:table_1}
\end{table}

We use Eq.~\ref{eq:phase_accumulated_integral} and the average gate fidelity to numerically find the ramp and gate times for the control pulse~\eqref{eq:smoothed_square_pulse} that produces a high-fidelity $z$-rotation about some target angle. The average gate fidelity is defined as~\cite{Pedersen2007}
\begin{equation}\label{eq:fidelity_equation}
\mathcal{F}=\frac{1}{m(m+1)}\left(\Tr\left[\tilde{U}\tilde{U}^{\dagger}\right]+\abs{\Tr\left[\mathcal{U}^{\dagger}\tilde{U}\right]}^2 \right),
\end{equation}
where $m$ is the dimension of both evolution operator $\tilde{U}$ and target operation $\mathcal{U}$. We find that using $\xi_0=\Delta E_{idle}=60~\text{kV~m}^{-1}$ and $\xi_f=\Delta_{op}=0.4~\text{kV~m}^{-1}$, i.e. the idling point is near the interface and the operating point is at the clock transition, a $\pi$ $z$-rotation can be generated in $T=70.35~\text{ns}$ with a ramp time $t_r=4.3~\text{ns}$ and $\varsigma=70$. The fidelity of this rotation in the absence of noise is 99.999\%. This is similar to the result for a $\pi$ $z$-gate shown in Ref.~\cite{Tosi2017b} (it is not exactly equal because we use a slightly different control pulse). Alternatively, using the same idling point but a different operating point closer to the donor~\cite{Simon2020}, $\Delta E_{op}=-12~\text{kV~m}^{-1}$, we can implement a $\pi$ $z$-rotation with the same fidelity (in the absence of noise) as before but with a much shorter gate time \rev{$T=23~\text{ns}$} ($t_r=0.9~\text{ns}$ and $\varsigma=1$). Figure~\hyperref[fig:pi fidelity, other fidelity, filter function]{\ref*{fig:pi fidelity, other fidelity, filter function}(a)} shows that with the same pulse parameters we can implement fast high-fidelity $z$-rotations by any arbitrary angle. 

Charge noise is the main source of decoherence in quantum devices based on isotopically purified silicon ($^{28}$Si), and it can be caused by nearby charge fluctuators~\cite{Paladino2014}. \rev{Other sources of noise, e.g., Johnson-Nyquist  noise and high-frequency noise due to voltage noise at the metallic gates,  are expected to be negligible or can be effectively suppressed via hardware modifications like inserting low-temperature attenuation along the high-frequency lines, which ensures the metal gates are well thermalized and substantially attenuates the noise of the room-temperature electronics~\cite{Tosi2017b}.}
Charge noise typically has a power spectral density that varies approximately as $1/f$ over a large range of frequencies $f$. In the flip-flop system, charge noise introduces electrical fluctuations that affect the control electric field $E_z(t)$. \rev{ The tunnel coupling $V_t$ can also be affected by overlap variations between the donor and interface wavefunctions due to fluctuations on the interface potential landscape, which can be caused by gate voltage noise or other sources of charge noise.} Owing to the large low-frequency component of the noise spectrum, a general approach for handling this type of noise influence on the system is to treat  \rev{the voltage noise and} the averaged collective effect of the nearby charge fluctuators as quasi-static perturbations, i.e. the noise is assumed constant during the gate time. Accordingly, we calculate the gate infidelity $1-\mathcal{F}$, Eq.~\eqref{eq:fidelity_equation}, of some of the gates reported above for different strengths of the electric field noise $\delta E_{z,\text{rms}}$ \rev{and a fixed tunnel coupling noise amplitude $\delta V_{t,\text{rms}}$. The latter is estimated from the simulation data for $V_t$ as a function of the top metal gate voltage $V_r$ presented in Fig.~2(g) of Ref.~\onlinecite{Tosi2017b}. We assume a 10~$\mu$V r.m.s noise~\cite{Tosi2017b,Dial2013} in $V_r$ to estimate $\delta V_{t,\text{rms}}$.
The upper part of Fig.~\hyperref[fig:pi fidelity, other fidelity, filter function]{\ref*{fig:pi fidelity, other fidelity, filter function}(b)} shows $\pi$ $z$-rotation infidelities averaged over the strength of the quasi-static electric field and tunnel coupling noises by sampling random perturbations $\delta E_{z}$  and $\delta V_t$ (linearly added to $\Delta E(t)$ and $V_t$, respectively) over uniform distributions with range $\sqrt{3}[-\delta E_{z,\text{rms}},\delta E_{z,\text{rms}}]$  and $\sqrt{3}[-\delta V_{t,\text{rms}},\delta V_{t,\text{rms}}]$. The average is taken over 200 samples for each value of $\delta E_{z,\text{rms}}$, ranging from 0.05 kV~m$^{-1}$ and 19.95 kV~m$^{-1}$, and 200 samples for the  value of $\delta V_{t,\text{rms}}$ in Table~\ref{tab:table_1}  associated to each $z$-rotation. The lower part of Fig.~\hyperref[fig:pi fidelity, other fidelity, filter function]{\ref*{fig:pi fidelity, other fidelity, filter function}(b)} presents the change in infidelity $\Delta_{1-\mathcal{F}}$ when only electric field noise is taken into account. This shows that the impact of the tunnel coupling noise on the gate infidelity is, on average, an order of magnitude lower than the estimated infidelity when only electric field noise is considered, e.g., if $1-\mathcal{F}$ is on the order of $10^{-4}$ with only electric field noise, then including the tunnel coupling noise in the calculation would modify $1-\mathcal{F}$ on the order of $10^{-5}$ or less.}
Now, in the particular case of the flip-flop system, Ref.~\cite{Tosi2017b} estimates that the r.m.s. amplitude of the quasistatic electric field noise affecting the system along the $z$-axis is $\sim 100~\text{V~m}^{-1}$. 
In Fig.~\hyperref[fig:pi fidelity, other fidelity, filter function]{\ref*{fig:pi fidelity, other fidelity, filter function}(b)}, the first curve $R_z(\pi)^{(\alpha)}$, which has the clock-transition as the operating point, presents a $\sim 99.95\%$ fidelity at $\delta E_{z,\text{rms}}=0.1~\text{kV~m}^{-1}$ \rev{and $\delta V_{t,\text{rms}}/2\pi=2.7~\text{MHz}$},  and a gate time of $T=70.35$~ns. The overall fidelity, however, can be bumped up by choosing an operating point even closer to the donor. For example, $R_z(\pi)^{(\beta)}$ in Fig.~\hyperref[fig:pi fidelity, other fidelity, filter function]{\ref*{fig:pi fidelity, other fidelity, filter function}(b)}  is generated by a pulse~\eqref{eq:smoothed_square_pulse} with an operating point closer to the donor $\xi_f=\Delta E_{op}=-12~\text{kV~m}^{-1}$ and has a $99.994\%$ fidelity under realistic noise amplitudes $\delta E_{z,\text{rms}}=0.1~\text{kV~m}^{-1}$  \rev{and $\delta V_{t,\text{rms}}/2\pi=2.7~\text{MHz}$}, and a much shorter gate time $T=23$~ns.  Fast high-fidelity gates can also be produced with a stronger magnetic field and an operating point closer to the donor, e.g. the third curve in Fig.~\hyperref[fig:pi fidelity, other fidelity, filter function]{\ref*{fig:pi fidelity, other fidelity, filter function}(b)} $R_z(\pi)^{(\gamma)}$ has $\Delta E_{op}=-20~\mathrm{kV~m}^{-1}$ as the operating point and a magnetic field $B_0=0.8~\mathrm{T}$; it has a \rev{$99.992\%$} fidelity and a gate time \rev{$T=12.36$}~ns at the noise amplitudes $\delta E_{z,\text{rms}}=0.1~\text{kV~m}^{-1}$  \rev{and $\delta V_{t,\text{rms}}/2\pi=3.3~\text{MHz}$}, which is much shorter than the gate with the clock transition as the operating point. Similarly, for a magnetic field $B_0=1.2~\mathrm{T}$ we predict $\pi$ $z$-rotations with a \rev{$99.993\%$} fidelity and a gate time \rev{$T=24.04$}~ns at the noise amplitudes $\delta E_{z,\text{rms}}=0.1~\text{kV~m}^{-1}$ \rev{and $\delta V_{t,\text{rms}}/2\pi=3.5~\text{MHz}$. Finally, for each $z$-rotation in Fig.~\hyperref[fig:pi fidelity, other fidelity, filter function]{\ref*{fig:pi fidelity, other fidelity, filter function}(b)}, we find that variations in the control pulse length of less than $\pm 0.1$~ns have a negligible effect on the fidelity, but variations greater than $\pm 0.2$~ns can reduce the fidelity by at least one order of magnitude (see Appendix~\ref{app:gate_time_over_undershoot} for more detail).} 

The use of an operating point closer to the donor results in faster and high-fidelity $z$-rotations because of the relative magnitude and shape of the flip-flop transition energy $\varepsilon_{ff}$ as it gets closer to the donor (see Fig.~\ref{fig:energy and transition frequency}). In this region, the magnitude of $\varepsilon_{ff}$ is larger than its value at the clock transition, which speeds up the rotation (see Eq.~\eqref{eq:phase_accumulated_integral}) and raises the qubit quality factor. Also, its slope ($\partial_{\Delta E}\varepsilon_{ff}$) decreases as it gets closer to the donor and the time spent near the ionization point is minimal, factors which combine to minimize the dephasing errors. \rev{Another advantage of using smooth pulses and $\Delta E_{op}$ closer to the donor is that fast high-fidelity $z$-rotations can be generated with rather low tunnel coupling values. In Appendix~\ref{app:gate_fidelity_z_weak_Vt}, we show numerical results demonstrating that with tunnel couplings of just a few GHz it is possible to generate fast high-fidelity $R_z$ gates in the presence of noise.} Moreover, without the need of a clock transition, there is more freedom to explore different sets of parameters that may lead to an overall better qubit performance. This has a direct impact on the relaxation time, since using an operating point closer to the donor increases the magnitudes of both $\varepsilon_0$ and $T_1$ considerably. For example, for $B_0=0.4$ T and $V_t=11.44$ GHz, using $\Delta E_{op}=-12~\mathrm{kV~m}^{-1}$ as operating point instead of the clock transition $\Delta E_{op}=0.4~\mathrm{kV~m}^{-1}$, increases the relaxation time five orders of magnitude.

\section{ $R_x(\phi)$ gates and effective Hamiltonian with oscillating driving}\label{sec:Rx_gates}
The implementation of an $x$-rotation about an arbitrary angle ($R_x(\phi)$) requires the use of an oscillating electric field to drive transitions between the flip-flop qubit states. The electric field, then, is given by $\Delta E(t)=\Delta E^{(\text{dc})}(t)+E^{(\text{ac})}(t)\cos(\omega t+\varphi)$, where $\Delta E^{(\text{dc})}(t)$ ($E^{(\text{ac})}(t)$) is the dc (ac) amplitude of the electric field. The use of an oscillating field incorporates the following energy term to the system Hamiltonian (Eq.~\eqref{eq:full_Hamiltonian}):
\begin{equation}\label{eq:Ac_component of the Hamiltonian}
    H_{orb}^{(\text{ac})}=-\frac{\varepsilon_{\text{ac}}}{2}\cos(\omega t+\varphi)\left(\sin\theta\,\sigma_{xi}+\cos\theta\,\sigma_{zi}\right),
\end{equation}
where $\varepsilon_{\text{ac}}=d\,e\,E^{(\text{ac})}/\hbar$. In the basis $\{\ket*{g~\widetilde{\uparrow\Downarrow}},\ket*{g~\widetilde{\downarrow\Uparrow}},\ket*{e~\widetilde{\uparrow\Downarrow}},\ket*{e~\widetilde{\downarrow\Uparrow}}\}$, the simplified Hamiltonian~\eqref{eq:H_matrix_form_spin_orbit_eigenbases} with ac driving has the following form 
\begin{widetext}
\begin{equation}\label{eq:H_AC_matrix_form_spin_orbit_eigenbases}
\tilde{H}^{(\text{ac})}\!=\!\left(
\begin{array}{cccc}
\frac{-(\varepsilon_0-\varepsilon_{s^{(-)}})}{2}+\Lambda(t)\cos\theta& 
 0 & 
  \Phi^{^{\!(+)}}\!\!(t)\sin\theta& 
 -\frac{1}{4}A\sin\theta \\
 0 & 
 \frac{-(\varepsilon_0+\varepsilon_{s^{(-)}})}{2}+\Lambda(t)\cos\theta & 
 -\frac{1}{4}A\sin\theta & 
  \Phi^{^{\!(-)}}\!\!(t)\sin\theta\\
 \Phi^{^{\!(+)}}\!\!(t)\sin\theta &
 -\frac{1}{4}A\sin\theta & 
 \frac{(\varepsilon_0+\varepsilon_{s^{(+)}})}{2}-\Lambda(t)\cos\theta & 
 0 \\
 -\frac{1}{4}A\sin\theta & 
 \Phi^{^{\!(-)}}\!\!(t)\sin\theta & 
 0 & 
 \frac{(\varepsilon_0-\varepsilon_{s^{(+)}})}{2}-\Lambda(t)\cos\theta\\
\end{array}
\right),
\end{equation}
\end{widetext}
where $\Lambda(t)=\tfrac{1}{8}A -\tfrac{1}{2}\varepsilon_{\text{ac}}\cos(\omega t+\varphi)$ and $\Phi^{^{\!(\pm)}}\!\!(t)=\frac{1}{8}(A\pm 2\Delta_{\varepsilon_z})-\frac{1}{2}\varepsilon_{\text{ac}}\cos(\omega t+\varphi) $.
Let $\delta\varepsilon$ be the smallest difference between diagonal energy levels from different diagonal blocks in $\tilde{H}^{(\text{ac})}$. Then, given that the non-oscillating elements and the oscillating amplitude $\varepsilon_{\text{ac}}$ in the off-diagonal blocks of the Hamiltonian~\eqref{eq:H_AC_matrix_form_spin_orbit_eigenbases} are  smaller than $\delta\varepsilon$, we can use the time-dependent SW (TDSW) transformation to find an analytical effective Hamiltonian in the qubit space. However, as further explained in Appendix~\ref{appendix:SW transformation}, the driving frequency at resonance is comparable in magnitude to the dominant energy scales in the Hamiltonian and, as a result, a system of differential equations must be solved in order to find the transformation matrix~\cite{Goldin2000a}. This is in contrast to other approaches found in the literature~\cite{Romhanyi2015,Theis2017} where the transformation matrix is found by solving a system of algebraic equations. 

\begin{figure*}[tbp]
\includegraphics[width=\linewidth]{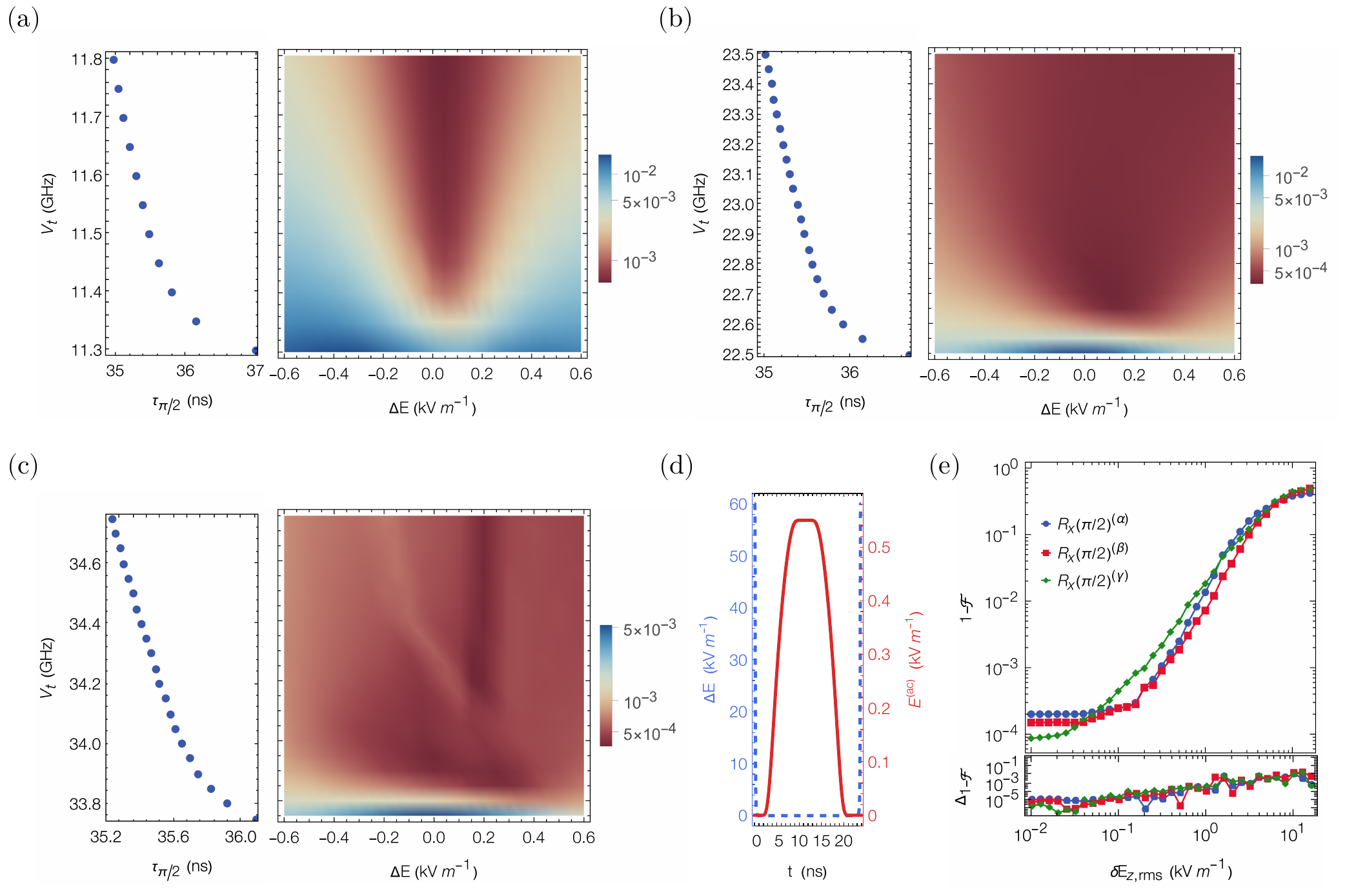}
\caption{Infidelity maps and average gate times of $R_x(\pi/2)$ gates for different magnetic field strengths: \textbf{(a)} $B_0=0.4$ T, \textbf{(b)} $B_0=0.8$ T, \textbf{(c)} $B_0=1.2$ T. The pulse operating point (tunnel coupling) is on the horizontal (vertical) axis of each infidelity map. We use two pulses, given by Eq.~\eqref{eq:smoothed_square_pulse}, one for the dc component (dashed blue curve in (d)) of the electric field and one for the ac component (solid red curve in (d)), the pulse parameters corresponding to the latter are labeled ``(ac)". All dc(ac) pulses start at $\xi_0=60~\mathrm{kV~m}^{-1}$($\xi_0^{\mathrm{(ac)}}=0~\mathrm{kV~m}^{-1}$) and have $\varsigma=1000$ and $t_r=1$. . The ac ramp time is given by $t_r^{\mathrm{(ac)}}=(T-2t_r)/\rho$, with $\rho=2.1$ being a scale factor, and all ac pulses have $\varsigma_{\text{ac}}=2$. Each point in the fidelity map is the result of averaging 100 samples taken from a uniformly distributed noise in the range $\sqrt{3}[-\delta E_{z,\text{rms}},\delta E_{z,\text{rms}}]$ with $\delta E_{z,\text{rms}}=100$ V m$^{-1}$. For the ac pulse, $\xi_f^{\mathrm{(ac)}}$ is easily obtained by solving Eq.~\eqref{eq:find_ac_max} with $R=0.23$. The plots to the left of each infidelity map depict the average gate time \rev{(horizontal axis)} under different values of the tunnel coupling \rev{(vertical axis)}. 
\rev{\textbf{(d)}  Electric field dc amplitude $\Delta E$ (dashed blue curve) and ac amplitude $E^{\mathrm{(ac)}}$ (solid red curve) pulse shapes used to implement $R_x(\pi/2)^{(\alpha)}$ in (e). 
\textbf{(e)} (Upper part) Infidelity of $\pi/2$ $x$-rotations as a function of various electric field noise strength values $\delta E_{z,\text{rms}}$ and fixed tunnel coupling noise $\delta V_{t,\text{rms}}$. The tunnel coupling noise and pulse parameters for each gate are given in Table~\ref{tab:table_2}. (Lower part) Change in infidelity $\Delta_{1-\mathcal{F}}$ between gate infidelity calculated with only electric field noise $\delta E_z$ and gate infidelity calculated with noise on both electric field $\delta E_z$ and tunnel coupling $\delta V_t$.
}} 
\label{fig:pi/2 x-gates infidelity}
\end{figure*}

Given that the coupling between the orbital ground and excited eigenstates is only non-negligible near the ionization point where $\theta\approx\pi/2$, in the TDSW transformation we neglect the Hamiltonian elements $\propto\cos\theta$. Moreover, the general solution to the system of differential equations that gives the TDSW transformation matrix $e^{S(t)}$ contains also terms $\propto e^{i\varepsilon_0 t}$, whose prefactors are set to zero owing to the requirement that $S(t)$ should be time-independent in the absence of oscillating driving ($\varepsilon_{ac}=0$). The effective ac-driven Hamiltonian in the qubit space is, therefore, given by:
\begin{equation}\label{eq:H_AC_flip-flop}
\begin{aligned}
    H_{ff}^{(\text{ac})}=&\frac{1}{2}\sum_{j=1}^3\left[(\Omega_{x,0}+\Omega_{x,j}\cos^j(\omega t+\varphi))\sigma_x \right.\\
   &\left.-(\Omega_{y,j-1}\cos^{j-1}(\omega t+\varphi))\sin(\omega t+\varphi)\sigma_y \right. \\
   &\left. -(\varepsilon_{ff,0}+\varepsilon_{ff,j}\cos^j(\omega t+\varphi))\sigma_z\right],
\end{aligned}    
\end{equation}
where $\Omega_{x,i}$, $\Omega_{y,i}$, and $\varepsilon_{ff,i}$ are given in Appendix~\ref{appebdix:Elements of the AC-driven H}.
We can go a step further and use Floquet perturbation theory~\cite{Shirley1965} to derive the effective Hamiltonian in the rotating frame and obtain analytical expressions for the Rabi frequency and resonance frequency. The Floquet method, in short, transforms a time-dependent Schr\"{o}dinger equation of a periodically driven finite-dimensional Hamiltonian $\mathcal{H}(t)$ into a time-independent Schr\"{o}dinger equation of an infinite-dimensional Floquet Hamiltonian $\mathcal{H}_F$ defined by~\cite{Shirley1965,Romhanyi2015}
\begin{equation}\label{eq: full_Floquet_matrix}
\mel{\alpha' m'}{\mathcal{H}_F}{\alpha m}=m\omega\delta_{\alpha' \!\alpha}\delta_{m'\! m}+\!\!\!\sum_{n=-\infty}^{\infty}\!\!\!\!\!\!\mel{\psi_{\alpha'}}{\mathcal{H}^{(n)}}{\psi_{\alpha}}\delta_{m'\!,n+m},
\end{equation}
where $\ket{\psi_{\alpha}}$ with $\alpha=1,\ldots,d_H$ ($d_H$ is the Hilbert space dimension) is an arbitrary basis of the Hilbert space and $\mathcal{H}^{(n)}$ are the Fourier components of the Hamiltonian, $\mathcal{H}(t)=\sum_{n=-\infty}^{\infty}\mathcal{H}^{(n)}e^{i n\omega t}$. In our case, the diagonal elements of the Floquet Hamiltonian $H_{F}$, obtained from applying the Floquet transformation to Eq.~\eqref{eq:H_AC_flip-flop}, form degenerate pairs when $\omega=-\varepsilon_{ff,0}-\varepsilon_{ff,2}/2$. For each of these pairs, the corresponding subspace is weakly coupled to the other diagonal elements and, therefore, it can be treated perturbatively using a time-independent SW transformation. To first-order, SW perturbation theory gives an effective $2\times2$ Floquet Hamiltonian
\begin{equation}\label{eq: effective_Floquet_Hamiltonian}
   \tilde{H}_F=
  \frac{1}{2} 
  \begin{pmatrix}
  -\Delta & \Omega_R e^{-i \varphi}\\
  \Omega_R e^{i \varphi} & \Delta
   \end{pmatrix},
\end{equation}
where $\Omega_R=\tfrac{1}{8}(4\Omega_{x,1}+3\Omega_{x,3}-4\Omega_{y,0}-\Omega_{y,2})$ and $\Delta=\tfrac{1}{2}(2\varepsilon_{ff,0}+\varepsilon_{ff,2}+2\omega)$. If $\Delta=0$, then the qubit is being driven at resonance and, therefore, the resonance and Rabi frequencies are $\omega_{\text{res}}=-\varepsilon_{ff,0}-\varepsilon_{ff,2}/2$ and $\Omega_{\text{res}}=\abs{\Omega_R}$, respectively. This result is exactly equal to the one obtained by neglecting the diagonal oscillating terms in the Hamiltonian and applying RWA. The effective Floquet Hamiltonian \eqref{eq: effective_Floquet_Hamiltonian} is, therefore, in a rotating frame defined by $U=\exp(-i\omega t\, \sigma_{z})$.  

\rev{The amplitude of the oscillating field, $E^{\text{ac}}$, should not be too large that it leads to leakage to higher states, nor should it be too small that the gate time becomes too long. We want a simple expression that can be used to tune  $E^{\text{ac}}$ to produce fast high-fidelity $x$-rotations. We can use the the ratio between the energy coupling logical states to higher states and the energy gap between those same states. This ratio should be $\ll 1$ to prevent leakage when using the smooth pulses introduced in the previous section. 
In the rotating frame, the flip-flop Hamiltonian with ac driving with electric field near or at the ionization point ($\Delta E=0$) presents small energy gaps between the logical states $\ket*{g~\widetilde{\uparrow\Downarrow}}$ and $\ket*{g~\widetilde{\downarrow\Uparrow}}$ and the higher state $\ket*{e~\widetilde{\downarrow\Uparrow}}$. On one hand, the coupling energy between $\ket*{g~\widetilde{\uparrow\Downarrow}}$ and $\ket*{e~\widetilde{\downarrow\Uparrow}}$  does not depend on $E^{\text{ac}}$ and is much smaller than their energy gap, and thus undesired transitions are highly unlikely. On the other hand,
the coupling energy between $\ket*{g~\widetilde{\downarrow\Uparrow}}$ and $\ket*{e~\widetilde{\downarrow\Uparrow}}$ does depend on $E^{\text{ac}}$ and can lead to leakage. We need analytic expressions for the $E^{\text{ac}}$-dependent coupling energy and the energy gap between  $\ket*{g~\widetilde{\downarrow\Uparrow}}$ and $\ket*{e~\widetilde{\downarrow\Uparrow}}$. We can get those analytic expressions using $\tilde{H}^{\text{(ac)}}$~\eqref{eq:H_AC_matrix_form_spin_orbit_eigenbases} in the rotating frame with  the approximation $\omega_{res}\approx \varepsilon_{ff}$. We use this approximation since the effective flip-flop transition energy $\varepsilon_{ff}$~\eqref{eq:effective_flip-flop_transition} is, by far, the dominant term in $\omega_{\text{res}}$ and, in contrast to the Rabi frequency, the correcting terms for the resonance frequency obtained with TDSW and Floquet theory are much smaller than $\varepsilon_{ff}$. After some simplifications, we find the following expression for the ratio between the $\ket*{g~\widetilde{\downarrow\Uparrow}}$-$\ket*{e~\widetilde{\downarrow\Uparrow}}$ coupling energy and energy gap:
\begin{equation}\label{eq:find_ac_max}
\frac{\varepsilon_{\text{ac}}\sin(\theta)}{4(\varepsilon_0-\varepsilon_{ff})}=R,
\end{equation}
where we set the ratio equal to $R$ with $0<R<0.5$. Depending on the system parameters, one can try different values for $R$ and use Eq.~\eqref{eq:find_ac_max} to find the value for $E^{\text{ac}}$ that would produce fast high-fidelity $x$-rotations.}

We find it convenient to use the same parameters $\{\varsigma=1000,\varsigma_{\text{ac}}=2, t_r=1, R=0.23 \}$ in all the calculations for $x$-rotations presented in Fig.~\hyperref[fig:pi/2 x-gates infidelity]{\ref*{fig:pi/2 x-gates infidelity}(a)-(c)} and Appendix~\ref{app:infidelity_maps}. The large value for $\varsigma$ suitably decreases the dc pulse slope in the region between the pulse inflection points and plateau of the control pulses, $\Delta E(t)$ and $E^{\mathrm{(ac)}}(t)$,that produce the desired $x$-rotation ($\varphi=0$). Hereafter, all the parameters for the ac pulse ($E^{\mathrm{(ac)}}$) are labeled ``(ac)". For the dc pulse, we use the same idle point that was used in the previous section $\Delta E_{idle}=60~\mathrm{kV~m}^{-1}$. The ac pulse always start at $E^{\mathrm{(ac)}}(t=0)=0~\mathrm{kV~m}^{-1}$ with a ramp time given by $t_r^{\mathrm{(ac)}}=(T-2t_r)/\rho$ with $\rho=2.1$, which ensures that the drive amplitude $E^{\mathrm{ac}}$ is non-zero only when the electron is at the operating point. We use Eq.~\eqref{eq:find_ac_max}, the analytical expressions for the resonance $\omega_{res}\approx \varepsilon_{ff}$ and Rabi $\Omega_{\mathrm{res}}$ frequencies, and the objective function
\begin{equation}\label{eq:gate time x rotation}
\Theta(T)=\abs{\text{mod}\left[\int_{t_r}^{T-t_r}\Omega_{\mathrm{res}}(t)\dd{t},2\pi\right]-\phi},
\end{equation}
to find the corresponding gate time $T$. Here, $\phi$ is the target rotation angle, and mod is the modulo operation. This procedure  gives a full set of parameters which produces high-fidelity $x$-rotations with the ac Hamiltonian in the dc eigenbasis.

\begin{table}[htb]
    \centering 
    \caption{\rev{System and pulse parameters for each infidelity curve in Fig.~\hyperref[fig:pi/2 x-gates infidelity]{\ref*{fig:pi/2 x-gates infidelity}(e)}. For each curve, the stated average gate fidelity $\mathcal{F}$ is calculated  assuming  electric field noise  $\delta E_{z,\text{rms}}=100~\mathrm{V~m}^{-1}$ and tunnel coupling noise $\delta V_{t,\text{rms}}$ indicated in this table. The ac ramp time is the same for each gate and is given by $t_r^{\text{(ac)}}=(T-2t_r)/2.1$.}}
    \vspace{1mm} 
\begin{tabular}{ |p{2.5cm}|| p{1.5cm}|p{1.5cm}|p{1.5cm}|}
 \hline
  & $R_x(\pi/2)^{(\alpha)}$ & $R_x(\pi/2)^{(\beta)}$ & $R_x(\pi/2)^{(\gamma)}$ \\
 \hline
 $B_0$~(T)   & 0.4 &  0.8 &  1.2\\
$ V_t/2\pi$~(GHz)   & 12.5   & 24.5 &  34.5 \\
$ \Delta E_{op}$~(kVm$^{-1}$)   & 0    & 0 &  1.5 \\
$ t_r$    & 1    & 1 &  1 \\
$ \varsigma$    & 1000   & 1000 &  1000 \\
$\varsigma^{\text{(ac)}}$  & 2  &  2  &  2  \\
$R$ &  0.38   &  0.4  &  0.4 \\
$E^{\text{(ac)}}$~(kVm$^{-1}$) & 0.55    &   0.94   &   0.62   \\
$\delta V_{t,\text{rms}}/2\pi$~(MHz) & 2.9 & 3.3 & 3.5  \\
$ T$~(ns)   & 23.86   & 23.42 & 24.23 \\
$ \mathcal{F}$~(\%)  & 99.98    & 99.98 &  99.96 \\
 \hline
\end{tabular}\label{tab:table_2}
\end{table}

Figures~\hyperref[fig:pi/2 x-gates infidelity]{\ref*{fig:pi/2 x-gates infidelity}(a)-(c)} show the infidelity maps for $\pi/2$ $x$-rotations for three different magnetic field strengths (0.4 T, 0.8 T, 1.2 T) commonly used in the laboratory and average gate times corresponding to different tunnel coupling values. The infidelities are averaged over the strength of a quasi-static noise by sampling a random perturbation $\delta E_{z}$, which is linearly added to $\Delta E(t)$, over a uniform distribution in the range $\sqrt{3}[-\delta E_{z,\text{rms}},\delta E_{z,\text{rms}}]$ with $\delta E_{z,\text{rms}}=100$ Vm$^{-1}$.  
\rev{In Figs.~\hyperref[fig:pi/2 x-gates infidelity]{\ref*{fig:pi/2 x-gates infidelity}(a)-(c)} we see that with an external magnetic field of ($i$) 0.4~T ($ii$) 0.8~T ($iii$) 1.2~T and $\Delta E=0$, a $V_t$ less than ($i$) 11.35~GHz ($ii$) 22.55~GHz ($iii$) 33.8~GHz leads to an average fidelity less than ($i$) 99.4\% ($ii$) 98.9\% ($iii$) 99.7\%  for noise level $\delta E_{z,\text{rms}}=100$ Vm$^{-1}$. 
In the upper part of Fig.~\hyperref[fig:pi/2 x-gates infidelity]{\ref*{fig:pi/2 x-gates infidelity}(e)} we present  the gate infidelities of three $\pi/2$  $x$-rotations, each with different magnetic field strength, for different strengths of the electric field noise $\delta E_{z,\text{rms}}$ and a fixed tunnel coupling noise amplitude $\delta V_{t,\text{rms}}$. The system and pulse parameters for these three gates are given in Table~\ref{tab:table_2}.  The control pulse shapes for $\Delta E(t)$ and $E^{\mathrm{(ac)}}(t)$ that are used to implement $R_x({\pi/2})^{(\alpha)}$ in Fig.~\hyperref[fig:pi/2 x-gates infidelity]{\ref*{fig:pi/2 x-gates infidelity}(e)} are shown in Fig.~\hyperref[fig:pi/2 x-gates infidelity]{\ref*{fig:pi/2 x-gates infidelity}(d)}. In contrast to the infidelity maps in  Fig.~\hyperref[fig:pi/2 x-gates infidelity]{\ref*{fig:pi/2 x-gates infidelity}(a)-(c)}, the average gate infidelity in \hyperref[fig:pi/2 x-gates infidelity]{(e)}  is obtained by sampling random perturbations $\delta E_z$ and $\delta V_t$ over uniform distributions with range $\sqrt{3}[-\delta E_{z,\text{rms}},\delta E_{z,\text{rms}}]$ and $\sqrt{3}[-\delta V_{t,\text{rms}},\delta V_{t,\text{rms}}]$, respectively. The infidelity average is taken over 200 samples for each value of $\delta E_{z,\text{rms}}$, and the same amount of samples for the value of $\delta V_{t,\text{rms}}$ given in Table~\ref{tab:table_2}. The lower part of Fig.~\hyperref[fig:pi/2 x-gates infidelity]{\ref*{fig:pi/2 x-gates infidelity}(e)} shows the change in infidelity $\Delta_{1-\mathcal{F}}$ that happens when only electric field noise is taken into account. Similarly to the $z$-rotation case in Sec~.\ref{sec:Rz_gates},  $\Delta_{1-\mathcal{F}}$ shows that including  tunnel coupling noise in the gate infidelity calculation produces a change that is at least an order of magnitude lower than the infidelity value obtained with only electric field noise. Lastly, for each $x$-rotation in Fig.~\hyperref[fig:pi/2 x-gates infidelity]{\ref*{fig:pi/2 x-gates infidelity}(e)} we find that shifts in the control pulse length of less than $\pm 1$~ns can at most reduce the gate fidelity by an order of magnitude (see Appendix~\ref{app:gate_time_over_undershoot} for further detail).}

The numerical results presented in Fig.~\ref{fig:pi/2 x-gates infidelity} show that our pulses can easily generate fast high-fidelity $x$-rotation for any magnetic field strength and a wide combination of tunnel coupling energies and electric field values. Moreover, in Appendix~\ref{app:infidelity_maps} we present an extended version of the maps presented in Fig.~\ref{fig:pi/2 x-gates infidelity}, which show that fast high-fidelity $x$-rotation can be implemented with large tunnel coupling energies and electric fields not necessarily close to the ionization point. The use of large tunnel coupling energies can also increase the relaxation time by a few orders of magnitude, even more if the best operating point is not near the ionization point like it is the case with $B_0=1.2$ T and $V_t>34$~GHz (see Fig.~\hyperref[fig:extended_infidelity_maps]{\ref*{fig:extended_infidelity_maps}(c)}).

\section{Conclusions}\label{sec: Conclusions}
We have presented control schemes to produce fast high-fidelity $z$- and $x$-rotations for flip-flop qubits in silicon.
Using both time-independent and time-dependent Schrieffer-Wolff transformations, and Floquet perturbation theories, we derived analytical expressions for the effective qubit Hamiltonian in the presence or absence of periodic driving. With these analytical expressions we numerically optimized the parameters of a modified Planck-taper window function such that it implements high-fidelity single-qubit gates in the shortest possible time. We proposed fast $z$- and $x$-rotations with fidelities around 99.99\% in the presence of realistic noise levels of $0.1~\mathrm{kV~m}^{-1}$, and gate times much shorter than previously reported. Moreover, since our method does not rely on sweet spots (clock transitions), we presented fast high-fidelity single-qubit gates with  magnetic fields stronger than what was previously proposed and closer to what is commonly used in the laboratory. Finally, the flexibility of our method allows the implementation of single-qubit gates with relaxation times and qubit quality factors five (one) order of magnitude larger than those corresponding to clock-transition-based $z$-rotations ($x$-rotations).

\section*{Acknowledgments}
We thank A. Morello for helpful discussions. This work is supported by the Army Research Office (W911NF-17-0287).

\appendix
\section{Time-dependent Schrieffer-Wolff perturbation theory}\label{appendix:SW transformation}

Before introducing the time-dependent Schrieffer-Wolff (TDSW) perturbation theory, we briefly review the time-independent version of it~\cite{Schrieffer1966,Winkler_2003}. 

Let us consider a general Hamiltonian $\mathcal{H}=\mathcal{H}_0+\mathcal{H}'$, where $\mathcal{H}_0$ is purely diagonal and $\mathcal{H}'=\mathcal{H}_1+\mathcal{H}_2$ is the perturbation. Assuming that the basis states of $\mathcal{H}$ are divided into two weakly interacting, energetically well-separated subspaces (diagonal blocks), then $\mathcal{H}_1$ is block-diagonal with zeroes as diagonal elements and $\mathcal{H}_2$ is strictly block-off-diagonal. The Schrieffer-Wolff transformation aims to decouple these two subspaces, transforming $\mathcal{H}$ into a block-diagonal Hamiltonian $\tilde{\mathcal{H}}$. In principle, $\tilde{\mathcal{H}}$ can be obtained via a unitary transformation: $\tilde{\mathcal{H}}=e^{-S}\mathcal{H}e^{S}$, where $S$ is a block-off-diagonal anti-Hermitian operator. In most of the cases, however, $S$ is not known and it must be constructed. This is done by first substituting $e^S$ in the unitary transformation with its series expansion, obtaining
\begin{equation}\label{eq:time-independent SW }
    \tilde{\mathcal{H}}=\sum_{j=0}^{\infty}\frac{1}{j!}\comm{\mathcal{H}_0+\mathcal{H}_1}{S}_{(j)}+\sum_{j=0}^{\infty}\frac{1}{j!}\comm{\mathcal{H}_2}{S}_{(j)},
\end{equation}
with $\comm*{\mathcal{H}}{S}_{(m+1)}=\comm*{\comm*{\mathcal{H}}{S}_{(m)}}{S}$ and $\comm*{\mathcal{H}}{S}_{(0)}=\mathcal{H}$. Since the block-off-diagonal unitary transformation $e^{S}$ must be close to unity due to the weakly interacting subspaces, then $S$ is small and can be expanded as a power series in the perturbation. Finally, each order of $S$ is determined successively by setting the block-off-diagonal part of $\tilde{\mathcal{H}}$ equal to zero and solving it order by order.

In TDSW, the block-off-diagonal anti-Hermitian operator $S$ is time dependent and, therefore, the unitary transformation that, in principle, can be used to obtain $\tilde{\mathcal{H}}(t)$ is now given by
\begin{equation}\label{eq:SW_time_dependent_unitary_transformation}
    \tilde{\mathcal{H}}(t)=e^{-S(t)}\mathcal{H}(t)e^{S(t)} + i\pdv{e^{-S(t)}}{t} e^{S(t)}.
\end{equation}
A time-dependent version of Eq.~\eqref{eq:time-independent SW } is obtained by plugging the series expansion of $e^{S}$ into Eq.~\eqref{eq:SW_time_dependent_unitary_transformation}:
\begin{widetext}
\begin{equation}\label{eq:time-dependent SW}
    \tilde{\mathcal{H}}=\sum_{j=0}^{\infty}\frac{1}{j!}\comm{\mathcal{H}_0+\mathcal{H}_1}{S}_{(j)}+\sum_{j=0}^{\infty}\frac{1}{j!}\comm{\mathcal{H}_2}{S}_{(j)}-i\sum_{j=0}^{\infty}\frac{1}{(j+1)!}\comm*{\dot{S}}{S}_{(j)},
\end{equation}
\end{widetext}
where $\dot{S}(t)=\partial_t S(t)$. Given that $S(t)$ is block-off-diagonal, the block-diagonal part $\tilde{\mathcal{H}}_{\text{diag}}$ of $\tilde{\mathcal{H}}$ contains the terms $\comm*{\mathcal{H}_0+\mathcal{H}_1}{S}_{(j)}$ with even $j$ and the terms $\comm*{\mathcal{H}_2}{S}_{(j)}$ and $\comm*{\dot{S}}{S}_{(j)}$ with odd $j$. The same goes for the block-off-diagonal part $\tilde{\mathcal{H}}_{\text{off}}$ but with odd $j$ instead of even $j$ and vice versa:
\begin{widetext}
\begin{align}
    \tilde{\mathcal{H}}_{\mathrm{off}}=&\sum_{j=0}^{\infty}\frac{1}{(2j+1)!}\comm{\mathcal{H}_0+\mathcal{H}_1}{S}_{(2j+1)}+\sum_{j=0}^{\infty}\frac{1}{(2j)!}\comm{\mathcal{H}_2}{S}_{(2j)}-i\sum_{j=0}^{\infty}\frac{1}{(2j+1)!}\comm*{\dot{S}}{S}_{(2j)},\\
    \tilde{\mathcal{H}}_{\mathrm{diag}}=&\sum_{j=0}^{\infty}\frac{1}{(2j)!}\comm{\mathcal{H}_0+\mathcal{H}_1}{S}_{(2j)}+\sum_{j=0}^{\infty}\frac{1}{(2j+1)!}\comm{\mathcal{H}_2}{S}_{(2j+1)}-i\sum_{j=0}^{\infty}\frac{1}{(2j+2)!}\comm*{\dot{S}}{S}_{(2j+1)}.\label{eq: appendix_diagonal_effective_Hamiltonian}
\end{align}
\end{widetext}

The expansion of $S=\sum_j S_j$ as a power series in the perturbation permits to solve $\tilde{\mathcal{H}}_{\text{off}}=0$ order by order. Here, $S_j$ is of $j$-th order in the perturbation. It is not immediately obvious, however, what order $\dot{S}_j$ is. Since the driving frequency $\omega$ is expected to characterize the time evolution of $S_j$, then we can assume that $\dot{S}_j\sim \omega S_j$~\cite{Romhanyi2015}. Now, in the particular case of the flip-flop qubit, for most values of $\Delta E$ the driving frequency $\omega$, the spin energy splitting $\varepsilon_s$, the hyperfine interaction $A$, and the driving amplitude energy $\varepsilon_{ac}$ are much smaller than the orbital splitting $\varepsilon_0$. However, around the ionization point, where the fastest $x(y)$-gates are obtained, $\omega\sim\varepsilon_s\sim\varepsilon_0$ and, therefore, $\omega$ cannot be treated as a perturbation. As a result, $\dot{S}_j$ and $S_j$ are both of $j$-th order in the perturbation.

The order-by-order expansion of $\tilde{\mathcal{H}}_{\mathrm{off}}=0$ gives a differential equation for each $S_j$ matrix operator. The first few equations are:
\begin{equation}\label{eq:appendix_set of differential equations Sj}
    \begin{aligned}
    \comm{\mathcal{H}_0}{S_1}&=-\mathcal{H}_2+i\dot{S}_1,\\
    \comm{\mathcal{H}_0}{S_2}&=-\comm{\mathcal{H}_1}{S_1}+i\dot{S}_2,\\
    \comm{\mathcal{H}_0}{S_3}&=-\comm{\mathcal{H}_1}{S_2}-\frac{1}{3}\comm{\mathcal{H}_2}{S_1}_{(2)}+i\dot{S}_3.\\
    \end{aligned}
\end{equation}
These equations, apart from determining the operator $S(t)$ in the transformation, can also be used to simplify Eq.~\eqref{eq: appendix_diagonal_effective_Hamiltonian}. The first few terms, then, that form the effective block-diagonal Hamiltonian $\tilde{\mathcal{H}}=\sum_j\tilde{\mathcal{H}}_j$ are:
\begin{equation}
    \begin{aligned}
    \tilde{\mathcal{H}}_0&=\mathcal{H}_0,\\
    \tilde{\mathcal{H}}_1&=\mathcal{H}_1,\\
    \tilde{\mathcal{H}}_2&=\frac{1}{2}\comm{\mathcal{H}_2}{S_1},\\
    \tilde{\mathcal{H}}_3&=\frac{1}{2}\comm{\mathcal{H}_2}{S_2},\\
    \tilde{\mathcal{H}}_4&=\frac{1}{2}\comm{\mathcal{H}_2}{S_3}-\frac{1}{4}\comm{\mathcal{H}_2}{S_1}_{(3)}.
    \end{aligned}
\end{equation}

\section{Analytical expressions for the elements of the ac-driven Hamiltonian}\label{appebdix:Elements of the AC-driven H}
The elements of the ac-driven Hamiltonian~\eqref{eq:H_AC_flip-flop} in the main text have the following form:
\begingroup
\allowdisplaybreaks
\begin{align*}
    \Omega_{x,0}=&\frac{A \sin ^2\theta }{8 \left(\varepsilon_0^2-\varepsilon_{s^{(-)}}^2\right)}\left(-\Delta_{\varepsilon_z} \varepsilon_{s^{(-)}}+\frac{A \left(2 \varepsilon_0^2-\varepsilon_{s^{(-)}}^2\right)}{2 \varepsilon_0} \right.\\
   &\left. -\frac{A^2 \varepsilon_0^2 (5 A \varepsilon_0-8 \Delta_{\varepsilon_z} \varepsilon_{s^{(-)}}) \sin ^2\theta}{32 \left(\varepsilon_0^2-\varepsilon_{s^{(-)}}^2\right)^2}\right)\\
   &+\frac{A \varepsilon_{\text{ac}}^2 \sin ^4\theta }{64 \left(\varepsilon_0^2-\varepsilon_{s^{(-)}}^2\right)^3}\Bigg( \Delta_{\varepsilon_z} \varepsilon_{s^{(-)}} \left(9 \varepsilon_0^2-\varepsilon_{s^{(-)}}^2\right) \\
   &-\frac{A \left(91 \varepsilon_0^4-126 \varepsilon_0^2 \varepsilon_{s^{(-)}}^2+51
   \varepsilon_{s^{(-)}}^4\right)}{16 \varepsilon_0}\Bigg),
\\
    \Omega_{x,1}=&-\frac{A \varepsilon_0 \varepsilon_{\text{ac}} \sin ^2\theta}{2 \left(\varepsilon_0^2-\varepsilon_{s^{(-)}}^2\right)}+\frac{A \varepsilon_{\text{ac}} \sin ^4\theta }{128 \varepsilon_0 \left(\varepsilon_0^2-\varepsilon_{s^{(-)}}^2\right)^3}\times\\
   &\bigg(A^2 \left(10 \varepsilon_0^2-\varepsilon_{s^{(-)}}^2\right) \left(\varepsilon_0^2+\varepsilon_{s^{(-)}}^2\right)\\
   &+\varepsilon_{\text{ac}}^2 \left(27
   \varepsilon_0^4-14 \varepsilon_0^2 \varepsilon_{s^{(-)}}^2+3 \varepsilon_{s^{(-)}}^4\right)\bigg),
 \\ 
    \Omega_{x,2}=&\frac{A \varepsilon_{\text{ac}}^2 \sin ^4\theta }{32
   \left(\varepsilon_0^2-\varepsilon_{s^{(-)}}^2\right)^3}\Bigg(-\Delta_{\varepsilon_z} \varepsilon_{s^{(-)}} \left(5 \varepsilon_0^2-2 \varepsilon_{s^{(-)}}^2\right)\\
   &+\frac{A \left(59 \varepsilon_0^4-102 \varepsilon_0^2 \varepsilon_{s^{(-)}}^2+27 \varepsilon_{s^{(-)}}^4\right)}{16 \varepsilon_0}\Bigg),
\\
    \Omega_{x,3}=&-\frac{A \varepsilon_{\text{ac}}^3 \sin ^4\theta \left(13 \varepsilon_0^4-34 \varepsilon_0^2 \varepsilon_{s^{(-)}}^2+5 \varepsilon_{s^{(-)}}^4\right)}{128 \varepsilon_0 \left(\varepsilon_0^2-\varepsilon_{s^{(-)}}^2\right)^3},
\\
    \Omega_{y,0}=&\frac{A^2 \varepsilon_{\text{ac}} \sin ^4\theta }{128 \left(\varepsilon_0^2-\varepsilon_{s^{(-)}}^2\right)^2}\Bigg(-9 \Delta_{\varepsilon_z} \varepsilon_{s^{(-)}}\\
    &+\frac{A \left(4 \varepsilon_0^4-9 \varepsilon_0^2
   \varepsilon_{s^{(-)}}^2+3 \varepsilon_{s^{(-)}}^4\right)}{\varepsilon_0 \left(\varepsilon_0^2-\varepsilon_{s^{(-)}}^2\right)}\Bigg),
\\
    \Omega_{y,1}=&\frac{A \varepsilon_{\text{ac}}^2 \sin ^4\theta }{32 \left(\varepsilon_0^2-\varepsilon_{s^{(-)}}^2\right)^3}\Bigg(-3 \Delta_{\varepsilon_z} \varepsilon_0^2 \varepsilon_{s^{(-)}}\\
   &-\frac{A \left(91 \varepsilon_0^4-126 \varepsilon_0^2 \varepsilon_{s^{(-)}}^2+51 \varepsilon_{s^{(-)}}^4\right)}{16 \varepsilon_0}\Bigg),
\\
    \Omega_{y,2}=&\frac{A \varepsilon_{\text{ac}}^3 \sin ^4\theta \left(27 \varepsilon_0^4-14 \varepsilon_0^2 \varepsilon_{s^{(-)}}^2+3 \varepsilon_{s^{(-)}}^4\right)}{128 \varepsilon_0 \left(\varepsilon_0^2-\varepsilon_{s^{(-)}}^2\right)^3},
\\
    \varepsilon_{ff,0}=&\varepsilon_{ff}+\frac{A \varepsilon_{\text{ac}}^2 \sin ^4\theta}{64 \left(\varepsilon_0^2-\varepsilon_{s^{(-)}}^2\right)^2}\Bigg(\frac{\Delta_{\varepsilon_z} \left(\varepsilon_0^2+\varepsilon_{s^{(-)}}^2\right)}{\varepsilon_0}\\
   &+\frac{A \varepsilon_{s^{(-)}} \left(9 \varepsilon_0^2-\varepsilon_{s^{(-)}}^2\right)}{\varepsilon_0^2-\varepsilon_{s^{(-)}}^2}\Bigg),
\\
    \varepsilon_{ff,1}=&\frac{\Delta_{\varepsilon_z} \varepsilon_{\text{ac}} \left(2 \varepsilon_0^2-\varepsilon_{s^{(-)}}^2\right) \sin ^2\theta}{4 \varepsilon_0 \left(\varepsilon_0^2-\varepsilon_{s^{(-)}}^2\right)}\\
    &+\frac{A^3 \varepsilon_{\text{ac}} \sin ^4\theta \left(4 \varepsilon_0^2 \varepsilon_{s^{(-)}}-3 \varepsilon_{s^{(-)}}^3\right)}{128
   \left(\varepsilon_0^2-\varepsilon_{s^{(-)}}^2\right)^3}\\
  & -\frac{\Delta_{\varepsilon_z} \varepsilon_{\text{ac}} \sin ^4\theta}{256 \varepsilon_0 \left(\varepsilon_0^2-\varepsilon_{s^{(-)}}^2\right)^3}\times\\
   &\Bigg(\varepsilon_{\text{ac}}^2
   \left(91 \varepsilon_0^4-126 \varepsilon_0^2 \varepsilon_{s^{(-)}}^2+51 \varepsilon_{s^{(-)}}^4\right)\\
   &+A^2 \bigg(28 \varepsilon_0^4-11 \varepsilon_0^2
   \varepsilon_{s^{(-)}}^2+6 \varepsilon_{s^{(-)}}^4-\frac{3 \varepsilon_{s^{(-)}}^6}{\varepsilon_0^2}\bigg)\Bigg),
\\
    \varepsilon_{ff,2}=&\frac{A \varepsilon_{\text{ac}}^2 \sin ^4\theta}{16 \left(\varepsilon_0^2-\varepsilon_{s^{(-)}}^2\right)^2}\Bigg(-\frac{A \varepsilon_{s^{(-)}} \left(5 \varepsilon_0^2-\varepsilon_{s^{(-)}}^2\right)}{2
   \left(\varepsilon_0^2-\varepsilon_{s^{(-)}}^2\right)}\\
   &-\frac{\Delta_{\varepsilon_z} \left(\varepsilon_0^2-2 \varepsilon_{s^{(-)}}^2\right)}{\varepsilon_0}\Bigg),
\\
    \varepsilon_{ff,3}=&\frac{\Delta_{\varepsilon_z} \varepsilon_{\text{ac}}^3 \sin ^4\theta \left(25 \varepsilon_0^2-13 \varepsilon_{s^{(-)}}^2\right)}{64 \varepsilon_0
   \left(\varepsilon_0^2-\varepsilon_{s^{(-)}}^2\right)^2}.
\end{align*}
\endgroup

\begin{widetext}
\section{Extended infidelity maps}\label{app:infidelity_maps}
Figure~\ref{fig:extended_infidelity_maps} shows extended versions of the infidelity maps and gate times for $\pi/2$ $x$-rotations shown in Fig.~\hyperref[fig:pi/2 x-gates infidelity]{\ref*{fig:pi/2 x-gates infidelity}(a-c)} of the main text.
\begin{figure*}[htb]
\includegraphics[width=\linewidth]{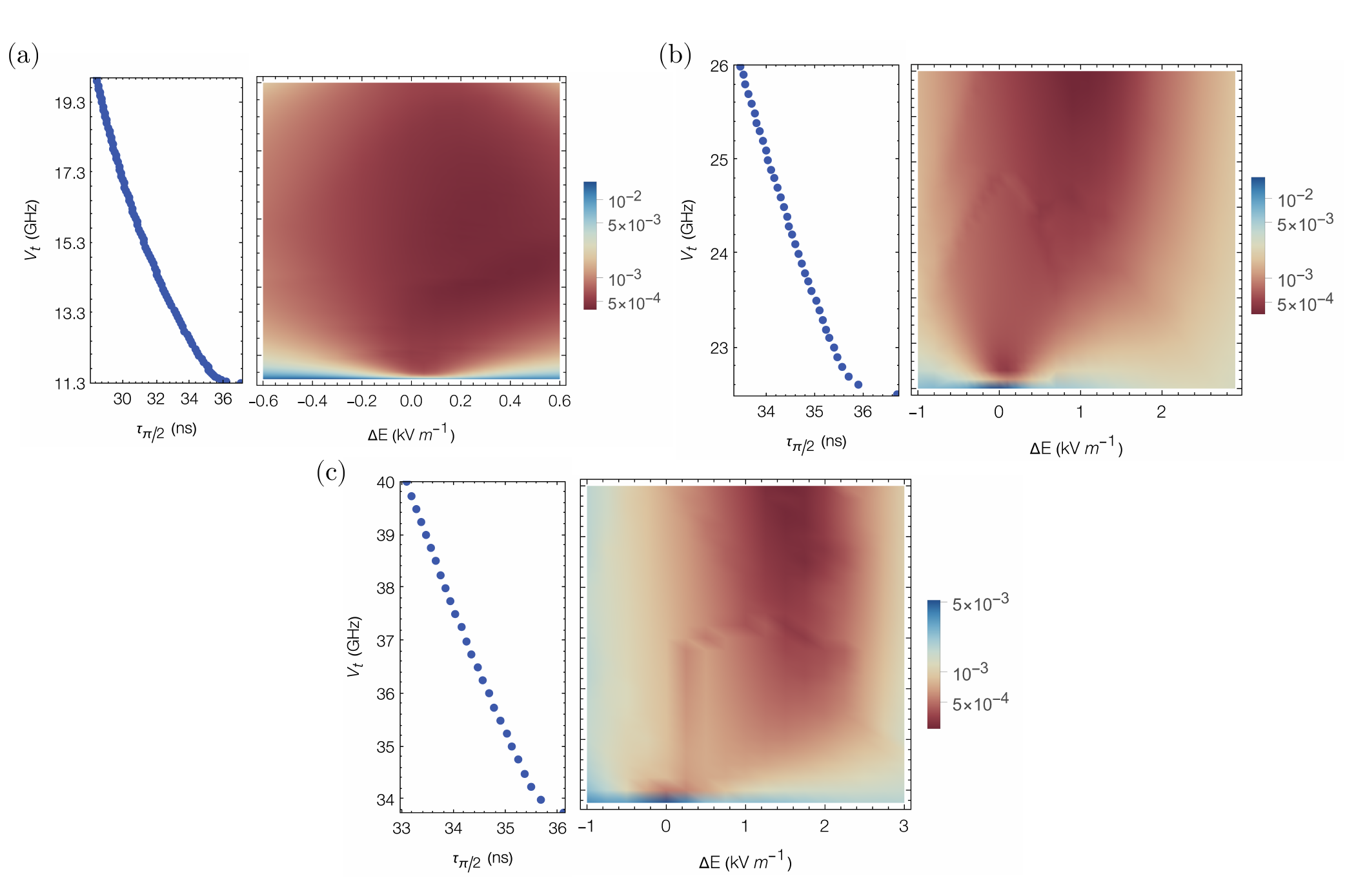}
\caption{Extended infidelity maps and average gate times corresponding to the maps in Fig.~\ref{fig:pi/2 x-gates infidelity} of the main text.}   
\label{fig:extended_infidelity_maps}
\end{figure*}

\section{\rev{Gate infidelity for $z$-rotations with weaker tunnel couplings}}\label{app:gate_fidelity_z_weak_Vt}
\rev{We show in Fig.~\ref{fig:min_Vt_Z_gates} gate infidelities and gate times for the same $z$-rotations from Fig.~\hyperref[fig:pi fidelity, other fidelity, filter function]{\ref*{fig:pi fidelity, other fidelity, filter function}(b)} in the main text. We calculate the gate infidelities and gate times using lower tunnel coupling values than those used in the main text. The gate infidelity is the result of averaging 100 samples for $\delta E_z$ taken from a uniform distribution in the range $\sqrt{3}[-\delta E_{z,\text{rms}},\delta E_{z,\text{rms}}]$ with $\delta E_{z,\text{rms}}=100$~Vm$^{-1}$. These results show that using an operating point $\Delta E_{op}$ closer to the donor instead of near to the ionization point generates fast high-fidelity $z$-rotations even for tunnel coupling values of a few GHz.}
\begin{figure}[htb]
\includegraphics[width=\linewidth]{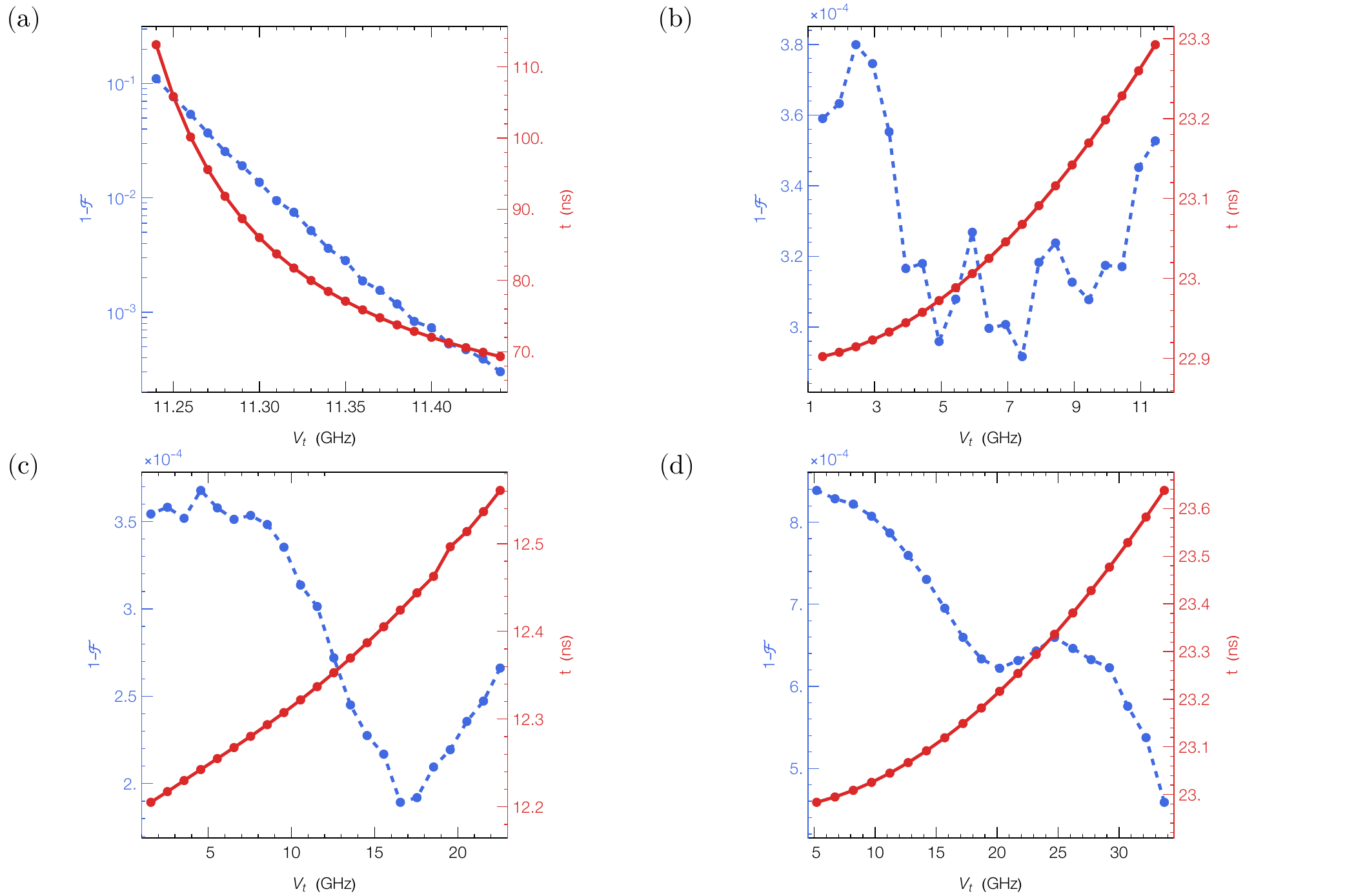}
\caption{\rev{Gate infidelities and gate times for the $z$-rotations \textbf{(a)} $R_z(\pi)^{(\alpha)}$, \textbf{(b)} $R_z(\pi)^{(\beta)}$, \textbf{(c)} $R_z(\pi)^{(\gamma)}$, \textbf{(d)} $R_z(\pi)^{(\delta)}$, presented in Fig.~\hyperref[fig:pi fidelity, other fidelity, filter function]{\ref*{fig:pi fidelity, other fidelity, filter function}(b)}. We use the same system and pulse parameters, except for the tunnel coupling, given in Table~\ref{tab:table_1}. }}   
\label{fig:min_Vt_Z_gates}
\end{figure}

\section{\rev{Gate infidelity sensitivity to pulse length perturbation} }\label{app:gate_time_over_undershoot}
\rev{Here we show the effect of pulse overshoot/undershoot on the infidelities of the gates presented in the main text. The gate infidelities shown in Fig.~\ref{fig:Delta_t_X_Z_gates} were obtained with the same system and pulse parameters of the $z$-rotations and $x$-rotations given by Tables~\ref{tab:table_1} and~\ref{tab:table_2} in the main text. In each case, in order to calculate the gate infidelity we average 100 sampled for $\delta E_z$ (electric field noise) taken from a uniform distribution in the range $\sqrt{3}[-\delta E_{z,\text{rms}},\delta E_{z,\text{rms}}]$ with $\delta E_{z,\text{rms}}=100$~Vm$^{-1}$. For $z$-rotations, variations in the pulse length of $\pm 1$~ns can lead to an infidelity increase between one and three orders of magnitude. For $x$-rotations, on the other hand, variations in the pulse length of $\pm 2$~ns can lead to an infidelity increase of  approximately one order of magnitude.}
\begin{figure*}[htb]
\includegraphics[width=\linewidth]{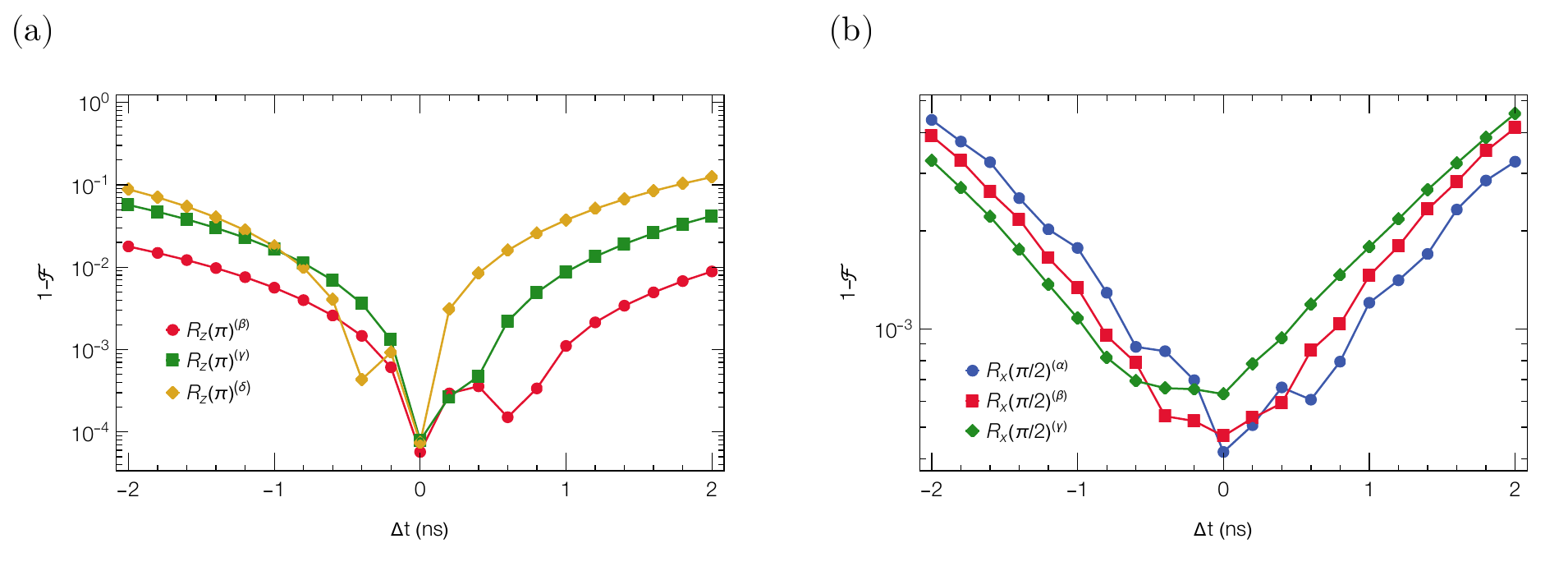}
\caption{\rev{Gate infidelities for \textbf{(a)} $z$-rotations presented in Fig.~\hyperref[fig:pi fidelity, other fidelity, filter function]{\ref*{fig:pi fidelity, other fidelity, filter function}(b)} and \textbf{(b)} $x$-rotations presented in Fig.~\hyperref[fig:pi/2 x-gates infidelity]{\ref*{fig:pi/2 x-gates infidelity}(e)}, obtained after perturbing their respective gate times by $\Delta t$.}}   
\label{fig:Delta_t_X_Z_gates}
\end{figure*}

\end{widetext}

\clearpage

\bibliography{library} 
\end{document}